\documentclass[12pt]{iopart}

\expandafter\let\csname equation*\endcsname\relax
\expandafter\let\csname endequation*\endcsname\relax
 
\usepackage{amsmath,comment,graphicx,color}

\newcommand{\vect}[1]{\mbox{\boldmath $#1$}}
\newcommand{\Bf}{\vect{B}_{\mathrm{f}}}
\newcommand{\BPM}{\vect{B}_{\mathrm{PM}}}

\begin{document}

\title[Calculation of permanent magnet arrangements for stellarators]{Calculation of permanent magnet arrangements for stellarators: A linear least-squares method}

\author{Matt Landreman$^1$ and Caoxiang Zhu$^2$}

\address{1. Institute for Research in Electronics and Applied Physics,
University of Maryland,
College Park, MD 20742, USA \\
2. Princeton Plasma Physics Laboratory,
Princeton, NJ 08543, USA}
\ead{mattland@umd.edu}
\vspace{10pt}

\begin{abstract}
A problem arising in several engineering areas is to design magnets outside a volume that produce a desired magnetic field inside it.
One instance of this problem is stellarator design, where it has recently been shown that permanent magnets can provide the required shaping of the magnetic field.
Here we demonstrate a robust and efficient algorithm 
REGCOIL\_PM
to calculate the spatial distribution of these permanent magnets. 
The procedure involves a small number of fixed-point iterations, with a linear least-squares problem solved at each step. The method exploits the Biot-Savart Law's exact linearity in magnetization density and approximate linearity in magnet size, for magnets far from the target region.
No constraint is placed on the direction of magnetization, so Halbach solutions are found naturally, 
and the magnitude of the magnetization can be made uniformly equal to a target value.
\end{abstract}

%
%
%
%
%

\section{Introduction}

Given a desired magnetic field in some region, what
arrangement of magnets outside the region can produce
this field? This problem has many applications,
including magnetic resonance imaging \cite{MRITurner, MRIPoole, MRIHidalgo}, particle accelerators \cite{Rossi,Russenschuck},
and stellarators for magnetic confinement of plasma \cite{NESCOIL,ONSET,COILOPT,REGCOIL,FOCUS}.
In stellarators, a three-dimensional magnetic field
must be carefully shaped in order to provide good confinement of charged particle trajectories and meet other physics objectives.
For all these applications, to design magnets one essentially needs
to invert the Biot-Savart law.
The Biot-Savart law provides a straightforward way to
compute the magnetic field $\vect{B}$ from known currents. However the inverse problem (given $\vect{B}$, determine currents) is ill-posed \cite{REGCOIL} in the sense that very different currents can give nearly the same $\vect{B}$.
Therefore there is room for creativity and innovation in formulating algorithms for these inverse problems such that the
magnet designs obtained are practical, and the computational cost is low.

While a variety of algorithms have been devised to compute
the shapes of electromagnetic coils \cite{NESCOIL,ONSET,COILOPT,REGCOIL,FOCUS},
in the area of stellarators
there is much less experience with algorithms for designing permanent magnets.
It has recently been demonstrated that stellarator fields
can be generated at least in part by permanent magnets \cite{HelanderPM,ZhuPerpendicular,Hammond,ZhuTopology}.
Electromagnetic coils are still needed to produce a net toroidal
field \cite{HelanderPM}, but permanent magnets can provide the remaining field.
Unlike electromagnets, permanent magnets would not require power supplies and would have greatly reduced need for cooling.
Other advantages of permanent magnets for stellarators include
the elimination of ripple due to discrete coils, and improved access to the plasma chamber for maintenance.
Disadvantages include the inability to turn off the field,
the possibility of demagnetization, and an upper limit on the achievable field strength.

In this paper we describe a new algorithm for the design of
permanent magnets. We call this method REGCOIL\_PM to highlight
its similarity to the REGCOIL method of electromagnetic coil design \cite{REGCOIL}.
REGCOIL\_PM differs from some previously proposed approaches for permanent magnet design \cite{HelanderPM,ZhuPerpendicular} in that no constraint is imposed
on the direction of the magnetization $\vect{M}$. Allowing the direction of $\vect{M}$ to be arbitrary is expected to reduce the necessary volume of permanent magnets by approximately a factor of 2 \cite{HelanderPM}. We do however aim to constraint the magnitude $M=|\vect{M}|$, making it everywhere uniform in the magnet region, since a technical upper limit exists, roughly $\mu_0 M \le 1.4$ T for present materials.

The REGCOIL\_PM algorithm is also formulated so as to be robust and fast.
These features are achieved by not formulating the task as a nonlinear optimization problem.
Numerical solution of nonlinear optimization problems can be fragile due to the existence of
multiple local minima, with the possibility of the solver stopping in a local minimum that is not the global optimum, resulting in sensitivity to the initial condition. Indeed, dependence of the solution on the initial guess was noted in \cite{ZhuTopology,Hammond}. Here instead we do not formulate the design problem as a nonlinear optimization problem.

The alternative approach here is motivated by the 
following expression for the magnetic field produced by a region of magnetization:
\begin{align}
    \BPM(\vect{r}) = \int_V d^3r'
    \frac{\mu_0}{4\pi |\vect{r} - \vect{r}'|^3} \left[ \frac{3(\vect{r}- \vect{r}')(\vect{r}- \vect{r}')\cdot\vect{M}(\vect{r}')}{|\vect{r} - \vect{r}'|^2}
    -\vect{M}(\vect{r}')\right].
    \label{eq:vol_integral}
\end{align}
Here, the integral is performed over a region $V$ containing the permanent magnets, and
$\BPM$
is the field at position $\vect{r}$ due to a magnetization (dipole moment density) $\vect{M}$ at position $\vect{r}'$.
Then if the magnet region $V$ is considered fixed,
(\ref{eq:vol_integral}) indicates that $\vect{B}(\vect{r})$ is a \emph{linear} function of $\vect{M}(\vect{r}')$. Regularization is required to make the problem well-posed, but Tikhonov-like  regularization can be introduced that preserves the linearity. Solving this linear problem for $\vect{M}$ is naturally faster and more robust than solving a nonlinear problem, and it naturally allows the direction of $\vect{M}$ to be arbitrary. 
(While $\vect{M}$ differs slightly from the zero-field magnetization $\vect{M}_0$, the difference is very small for rare-Earth magnets and so will be neglected here. If desired, the material's relationship $\vect{M}(\vect{B})$ could be inverted at each point to obtain $\vect{M}_0$.)

\begin{figure}
  \centering
  \includegraphics[width=4in]{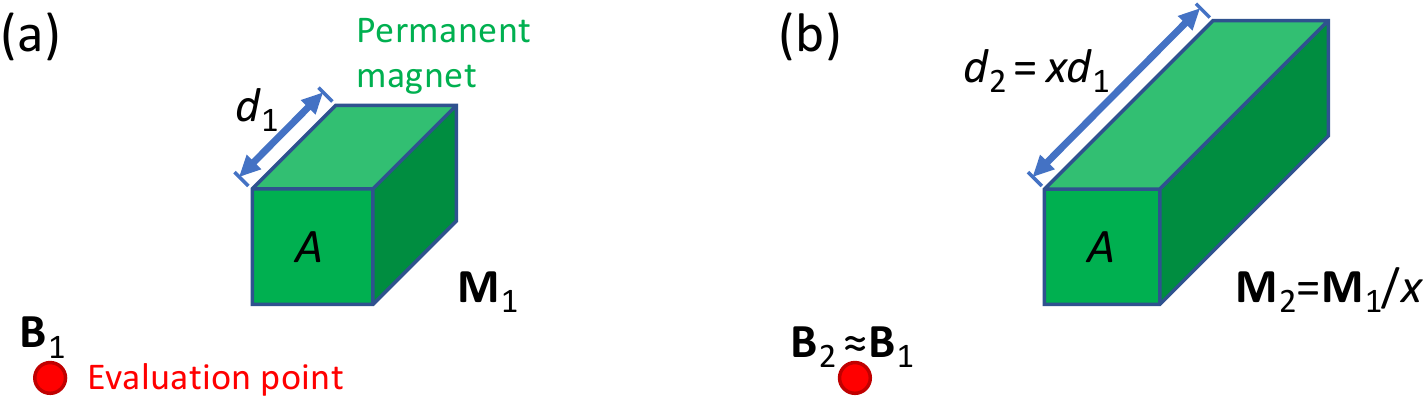}
  \caption{
Physical picture of the fixed-point iteration in the REGCOIL\_PM algorithm used to achieve a uniform magnetization $M=|\vect{M}|$.
(a) Suppose a region of magnetization $\vect{M}_1$ produces a given field $\vect{B}_1$ at a distant evaluation region. (b) If one dimension of the magnet is scaled by some factor $x$ and the magnetization is scaled by $1/x$, the change to $\vect{B}$ in the evaluation region is small.
  }
\label{fig:basicIdea}
\end{figure}

However the solution of the linear problem at fixed $V$ will have nonuniform $M = |\vect{M}|$, whereas it is preferable to have a solution with $M$ uniformly equal to the limit of the magnet material. Uniformity can be achieved by noting that the volume integral in (\ref{eq:vol_integral}) makes $\BPM$ approximately linear in the magnet thickness. The complicated dependence on $\vect{r}-\vect{r}'$ in (\ref{eq:vol_integral}) spoils this linearity, but if the magnets are not extremely close to the evaluation region, the nonlinearity will be weak. Thus, we can approximately correct the nonuniformity in $M$ by adjusting the magnet thickness. If $M$ is too large by a factor of two in a given region of $V$, doubling the thickness of $V$ in this region (at fixed $\vect{B}$) will result in a lowering of $M$ by approximately the same factor of two. This relationship is illustrated in figure \ref{fig:basicIdea}. While $M$ will not be exactly uniform after this change to the shape of $V$, the procedure can be iterated to improve the uniformity. As we will demonstrate, the number of iterations  required can be quite small. While a Newton-type iteration could also be used to account for the nonlinearity, such a method would require either analytic derivatives or finite-difference derivatives, increasing the computational cost. We will show that the simpler Picard fixed-point iteration is stable and sufficient in practice.

A serious experimental design for a permanent magnet stellarator requires a detailed geometry model with many individual magnet pieces, as in \cite{Hammond}. Here instead we will make a crude approximation that the magnetization fills a single region with smooth curved boundaries. This smooth model roughly approximates a large number of individual magnets, but we acknowledge this approximation is likely insufficient for a serious experimental design. We make this smooth approximation here for several reasons. The primary reason is that it allows reuse of a significant amount of the REGCOIL code \cite{REGCOIL}. This made it possible to try out the idea rapidly. Second, the smooth model is numerically convenient, since it allows us to evaluate integrals with spectral accuracy using uniform grids in periodic coordinates. Having demonstrated the REGCOIL\_PM algorithm in this paper, it should be straightforward to apply it in the future to more realistic geometry models with discrete magnets.

In the remainder of this paper, this REGCOIL\_PM algorithm is defined in greater detail and demonstrated for several problems. Section \ref{sec:formulation}
gives further detail about the mathematical formulation.
Aspects of the discretization and numerical solution are discussed in section \ref{sec:numerical}, using the NCSX stellarator as an example.
In section \ref{sec:benchmark} it is shown that our implementation reproduces the analytic solution for a Halbach cylinder. The NCSX example is developed and analyzed further 
in section \ref{sec:NCSX},
and we conclude in section \ref{sec:conclusions}.


\section{Mathematical formulation}
\label{sec:formulation}


We consider the common 2-stage approach to stellarator design.
In the first stage, the parameter space for optimization is the 
space of toroidal plasma boundary shapes,
and the objective function is a combination of physics figures of merit
for the plasma inside this boundary.
In the second stage, the shapes of magnets are optimized to
produce the plasma boundary shape resulting from the first stage.
Our goal in this paper is to solve the stage-2 problem.
If the stage-2 problem can be \emph{approximately} solved quickly,
the solution can be incorporated into the stage-1 objective function
to penalize magnet complexity \cite{Pomphrey2001}.
In this way, the stage-1 optimization can be made to find plasma
configurations that can be supported by magnets of low complexity,
and a more detailed and computationally demanding stage-2 calculation
can be done for the final magnet design.

We thus focus on the problem of finding a permanent magnet arrangement to produce
a desired plasma boundary surface $S$. To state this problem precisely,
first consider that $S$ must be a magnetic surface, so we wish to make
$\vect{B}\cdot\vect{n} \approx 0$ everywhere on $S$. (Matching the normal component is sufficient to ensure that the full vector $\vect{B}$
coincides with the target field everywhere inside $S$.)
We then use the linearity of magnetostatics to write $\vect{B} = \BPM + \Bf$ where $\BPM$ is the magnetic field (\ref{eq:vol_integral}) produced by the permanent magnets and $\Bf$ is
the field produced by currents that are ``fixed'' during the permanent magnet design.
The quantity $\Bf$ represents contributions from the electromagnets
and from current in the plasma, if there is any. Our goal can then be stated as achieving $f_B \approx 0$ where
\begin{align}
    f_B = \int_S \left[\left( \BPM + \Bf \right)\cdot \vect{n}\right]^2 d^2x,
\end{align}
an integral over the surface $S$ of the squared normal component of the field.

\begin{figure}
  \centering
  \includegraphics[width=6.1in]{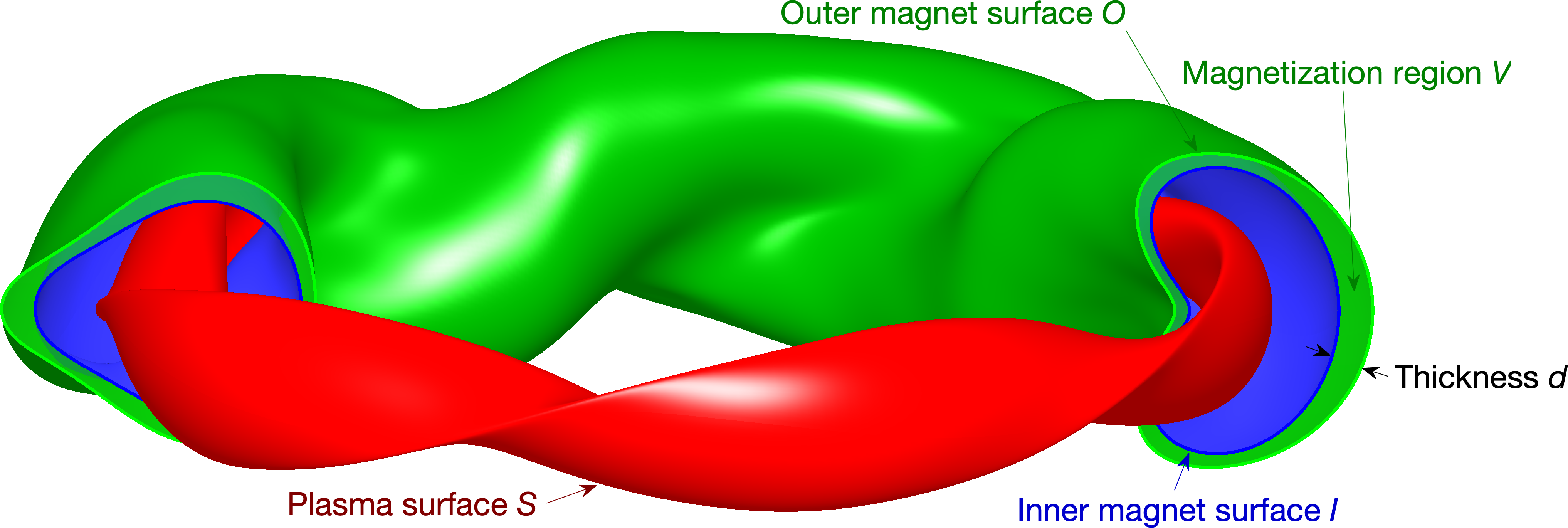}
  \caption{
Definitions of geometric quantities.
  }
\label{fig:geometryIllustration}
\end{figure}

The problem of finding $\vect{M}$ such that $f_B\approx 0$ is ill-posed, for an infinite number of widely different magnet distributions can produce nearly identical $\vect{B}$ on and inside $S$. 
Here, we will make the problem well-posed in two steps,
first constraining the magnet location and then adding a regularization term.
In the first step, we restrict the permanent magnets to lie within a volume $V(d)$
with some thickness parameter $d$. In this paper we will take $V$ to be bounded by two toroidal surfaces, a fixed inner surface $I$ and a variable outer surface $O$, both linking the plasma surface $S$, with $d$ a function on $I$ that measures the distance to $O$ (figure \ref{fig:geometryIllustration}). 
Specifically, we choose $V$ to be the range of position vectors 
\begin{align}
\label{eq:positionVector}
    \vect{r}'(s,\theta,\zeta)
    =\vect{r}_I(\theta,\zeta) + \sigma \, s \, d(\theta,\zeta)\vect{n}(\theta,\zeta)
\end{align}
 with $\theta \in [0,2\pi)$, $\zeta \in [0, 2\pi)$, and $s \in [0,1]$. Here, $\vect{r}_I(\theta,\zeta)$ is the position vector on the fixed inner surface $I$ of the magnet region, $\theta$ and $\zeta$ are any poloidal and toroidal angles, $d(\theta,\zeta)$ is a thickness function that will be varied, $\vect{n} = \vect{N}/|\vect{N}|$ is a unit normal of the inner surface, 
 \begin{align}
 \label{eq:Ndef}
     \vect{N} = \frac{\partial \vect{r}_I}{\partial\zeta} \times \frac{\partial \vect{r}_I}{\partial\theta}
 \end{align}
is a non-unit-length normal vector, and $\sigma = \pm 1$ is a constant chosen so the outer boundary $O$ of $V$ at $s=1$ is outside the inner surface $I$.
As long as $d$ does not exceed a $(\theta,\zeta)$-dependent threshold related to the curvature of $I$, the map $\vect{r}'(s,\theta,\zeta)$ in (\ref{eq:positionVector}) is invertible.
This continuous model for $V$ is an approximation to a set of many small discrete magnets.
Other choices for $V$ are possible, such as a set of $n$ discrete hexahedral volumes, with $d$ a vector of $n$ numbers giving the thickness of each region. However $V(d)$ is defined, we will take $d$ to be unknown, to be determined during the permanent magnet design. Restricting the permanent magnet location to a parameterized volume $V(d)$ is a reasonable reflection of practical engineering considerations: we expect the permanent magnets should be as close to the plasma as possible, limited by the vacuum vessel and any other components that may need to lie in between, but the thickness of the permanent magnet region depends on the specific plasma configuration.

However, restricting the permanent magnets to be located in
a specific $V(d)$ like (\ref{eq:positionVector}) does not fully eliminate the ill-posedness in the problem $f_B\approx 0$. 
Consider that near any point a finite distance from $S$ we could add two oppositely directed magnets with substantial magnetization, and if these magnets were sufficiently small and close to each other, the change to $\vect{B}$ on $S$ would be negligible. This type of ill-posedness is further discussed in \cite{REGCOIL}.
To arrive at a well-posed problem we therefore introduce a regularization term. The most convenient term to introduce is
\begin{align}
\label{eq:regularization}
    f_M = \int_V |\vect{M}|^2 \, w(\theta,\zeta) \, \,d(\theta,\zeta) \, d^3x,
\end{align}
a weighted volume integral over the permanent magnet region of the squared magnetization density. 
Here $w(\theta,\zeta)$ is an optional user-supplied weight function that can be used to exclude magnets from certain regions, such as where ports are to be placed. The appearance of the magnet thickness $d(\theta,\zeta)$
in (\ref{eq:regularization}) is motivated by the fixed-point iteration that will be explained shortly.
The form of (\ref{eq:regularization}) is essentially Tikhonov regularization, but with a physically meaningful weighting. (A similar regularization term without the weighting factors was proposed in appendix B of \cite{ZhuTopology}.) We can now define a combined objective function
\begin{align}
    f = f_B + \lambda f_M,
\end{align}
where $\lambda$ is a positive scalar parameter controlling the amount of regularization. Small values of $\lambda$ correspond to precisely making the target magnetic field (very small $f_B$) at the expense of more complicated permanent magnet structures, while large values of $\lambda$ yield simplified magnet structures at the expense of magnetic field inaccuracies (larger $f_B)$.
The problem of finding $\vect{M}$ that minimizes $f$ at fixed $V$ has the form of a linear least-squares problem.

We can now define the fixed-point iteration to make $M$ uniform. We constrain $\vect{M}$ to be independent of the radial coordinate $s$, so $\vect{M}=\vect{M}(\theta,\zeta)$.
The physical picture in figure \ref{fig:basicIdea} can then be expressed as $d_1(\theta,\zeta) M_1(\theta,\zeta) \approx d_2(\theta,\zeta) M_2(\theta,\zeta)$, where the subscripts 1 and 2 refer to a pair of magnet configurations such as panels (a)-(b) of figure \ref{fig:basicIdea}. If we desire for $M_2$ to equal a uniform target value $M_t$, then the appropriate update rule for $d$ is
\begin{align}
\label{eq:d_update}
    d_{j+1} = d_j \frac{M_j}{M_t}.
\end{align}

The factor of $d$ in (\ref{eq:regularization}) can now be explained. The iteration (\ref{eq:d_update}) preserves the product $Md$. In (\ref{eq:regularization}), the volume integral contains an explicit $M^2d$ factor as well as an implicit $d$ factor through the thickness of the integration region $V$. Therefore (\ref{eq:regularization}) is approximately constant during the iterations (\ref{eq:d_update}). The explicit $d$ factor in (\ref{eq:regularization}) is not critical, but it is convenient since a ``good'' value of $\lambda$ for the first iteration is likely to be a good value also for the final iteration.

The REGCOIL\_PM method can now be summarized. First, $d(\theta,\zeta)$ is initialized to a uniform thickness $d_1$, and choices of $\lambda$ and $M_t$ are fixed. Then the least-squares problem of minimizing $f$ (at fixed $d$) is solved for $\vect{M}_1$. An updated thickness $d_2(\theta,\zeta)$ is computed from (\ref{eq:d_update}). Using the new $V$ derived from $d_2$ and (\ref{eq:positionVector}), the least-squares problem is solved again to yield $\vect{M}_2$. An updated thickness $d_3(\theta,\zeta)$ is computed from (\ref{eq:d_update}), and the process is repeated until successive iterates are sufficiently close to each other.

To evaluate the volume integral in (\ref{eq:regularization}), we note
$\int_V = \int_0^{2\pi}d\theta \int_0^{2\pi}d\zeta \int_0^1ds |\sqrt{g}|$ where the Jacobian derived from (\ref{eq:positionVector}) is
\begin{align}
\label{eq:Jacobian}
    \sqrt{g}=-\sigma dN\left[1 - 2 \sigma s d H + (LQ-P^2) s^2 d^2 /N^2 
    \right],
\end{align}
where $N=|\vect{N}|$, $H=(LG+QE-2PF)/(2N^2)$ is the mean curvature of the inner surface $I$, and
\begin{align}
&    E = \frac{\partial \vect{r}_I}{\partial\theta} \cdot \frac{\partial \vect{r}_I}{\partial\theta},
    \hspace{0.5in}
    F = \frac{\partial \vect{r}_I}{\partial\theta} \cdot \frac{\partial \vect{r}_I}{\partial\zeta},
    \hspace{0.5in}
    G = \frac{\partial \vect{r}_I}{\partial\zeta} \cdot \frac{\partial \vect{r}_I}{\partial\zeta},
    \\
&
L = \vect{n}\cdot\frac{\partial^2 \vect{r}_I}{\partial\theta^2},
\hspace{0.5in}
P = \vect{n}\cdot\frac{\partial^2 \vect{r}_I}{\partial\theta\partial\zeta},
\hspace{0.5in}
Q = \vect{n}\cdot\frac{\partial^2 \vect{r}_I}{\partial\zeta^2}.
\nonumber \end{align}
This result is derived in  \ref{apx:Jacobian}.


\section{Numerical solution}
\label{sec:numerical}

We now discuss the discretization and numerical solution of the
equations of the previous section.
The source code for the numerical implementation used here
is available online at  \cite{zenodoCode},
and data for the figures and benchmarks is available at \cite{zenodoData}.


\subsection{Discretization}

The magnetization vector is first written as a finite sum of basis functions
\begin{align}
\label{eq:finiteSum}
    \vect{M}(\theta,\zeta) = \sum_{j=1}^{j_{\max}} \sum_{k=1}^3 M_{j,k} p_j(\theta,\zeta) \vect{e}_k(\zeta),
\end{align}
where
\begin{align}
\label{eq:basisFunctions}
    p_j(\theta,\zeta) = \begin{pmatrix} \sin\\ \cos \end{pmatrix}_j
    \left( m_j \theta - n_j n_{fp} \zeta\right)
\end{align}
are angular basis functions, and $\vect{e}_k$ for $k=1,2,3$ are the unit vectors for cylindrical coordinates, $\vect{e}_R$, $\vect{e}_\phi$, $\vect{e}_Z$.
The notation in (\ref{eq:basisFunctions}) means that either $\sin$ or $\cos$ is chosen for basis function $j$. The number of identical field periods is denoted $n_{fp}$. The integers $m_j$ range from 0 to $m_{\max}$, and $n_j$ ranges from 0 to $n_{\max}$ (for $m_j=0$) or $-n_{\max}$ to $n_{\max}$ (for $m_j>0$).
We take $\zeta$ to be equal to the standard toroidal angle $\phi$ on the inner surface,
so  (\ref{eq:positionVector}) implies that $\zeta$ is \emph{not} generally the standard toroidal angle off of $I$. The vectors $\vect{e}_k(\zeta)$ are evaluated at the point on $I$ with the given $\zeta$, meaning that they generally differ from the cylindrical basis vectors at any point off $I$ where $\vect{M}$ is evaluated. The reason for this choice is so the Cartesian components of $\vect{M}$ remain constant as you move in the normal direction from $I$, reflecting a reasonable engineering constraint.

In the common case of stellarator symmetry, only the $\sin(m_j\theta-n_j\zeta)$ basis functions need to be included in (\ref{eq:finiteSum}) for the $\vect{e}_R$ terms, and only the  $\cos(m_j\theta-n_j\zeta)$ basis functions need to be included for the $\vect{e}_\phi$ and $\vect{e}_Z$ terms.

If one wished to allow $\vect{M}$ to vary with $s$, a sum over basis functions in $s$ (such as polynomials) could be included in (\ref{eq:finiteSum}). Our numerical implementation allows this possibility.
However if this $s$ dependence is allowed, it is hard to see how to achieve a uniform $M$ by the fixed-point iteration proposed here. Therefore for all results in this paper we do not include $s$ dependence in (\ref{eq:finiteSum}).

The objective function $f_B$ involves an integral over the plasma boundary surface $S$. This integral is written as a discrete sum using a uniformly spaced grid of $N_\theta$ points in $\theta$ and a uniform grid of $N_\zeta$ points in $\zeta$, with $\theta$ and $\zeta$ angles on $S$ in this case.
Moreover, to evaluate $f_B$ and $f_M$, integrals over the magnetization volume are required. These integrals are written as discrete sums using a uniform grid of $N_{\theta'}$ points in $\theta$, a uniform grid of $N_{\zeta'}$ points in $\zeta$, and a Gauss-Legendre grid of $N_s$ points in $s$.
The magnet thickness $d(\theta,\zeta)$ is stored on the same $N_{\theta'}\times N_{\zeta'}$ discrete grid points.

It could be reasonable to take the independent variables as the components of $\vect{M}$ on the discrete $N_{\theta'}\times N_{\zeta'}$ grid points, instead of using the Fourier amplitudes $M_{j,k}$ in (\ref{eq:finiteSum}).
We choose the Fourier approach here due to two advantages. First, it is convenient for imposing stellarator symmetry, which reduces the number of degrees of freedom by a factor of two. Second, it allows the volume integrals in $f_B$ and $f_M$ to be evaluated at higher spatial resolution without increasing the number of degrees of freedom for the least-squares solution, which is numerically efficient in practice.
In the Fourier representation used here, one should choose $N_{\theta'} \ge 2 m_{\max}+1$ and $N_{\zeta'}/n_{fp} \ge 2 n_{\max} + 1$ so there are at least as many degrees of freedom in the grid as in Fourier space. Otherwise $f_M$ does not fully regularize every Fourier mode.

With $\vect{M}$ represented by the finite sum (\ref{eq:finiteSum}), and the integrals over $S$ and the magnetization volume in $f_B$ and $f_M$ approximated by finite sums as described above, minimization of $f$
now has the form of a finite linear-least-squares problem. Such problems
can be solved by standard methods such as the normal equations, $QR$ decomposition, or singular value decomposition.


\subsection{Least-squares problem}
\label{sec:leastSquares}

\begin{figure}
  \centering
  \includegraphics[width=6.1in]{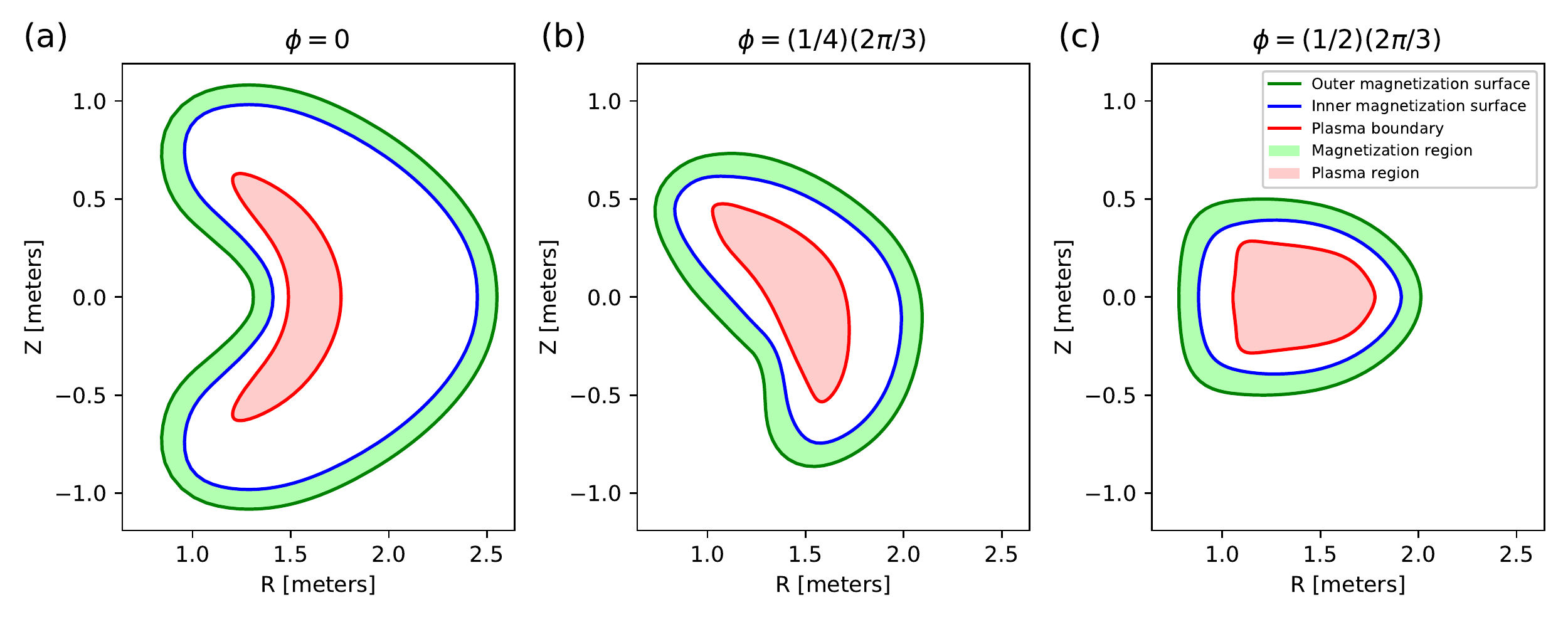}
  \caption{
Geometry for the discussion of regularization and resolution parameters in section \ref{sec:leastSquares}.
Here, $\phi$ is the standard toroidal angle, which corresponds to $\zeta$ on the inner surface $I$ but not off of $I$.
  }
\label{fig:XSections_uniformD}
\end{figure}

Before demonstrating the entire REGCOIL\_PM algorithm,
it is valuable to first examine the behavior of the least-squares solution at fixed $d$, without the fixed-point iteration. For this discussion
we will use the geometry shown in figure \ref{fig:XSections_uniformD}.
This figure shows slices through the geometry at constant $\phi$, where $\phi$ is the standard toroidal angle, coinciding with $\zeta$ only on the inner magnet surface $I$.
The plasma geometry is that of the c09r00 version of NCSX, a free-boundary equilibrium computed using infinitesmally thin approximations of the 18 discrete modular coils. For this paper we will use the c09r00 boundary shape but neglect plasma currents. The contribution to $\Bf$ from plasma current could be computed using the same virtual-casing method \cite{ROSE} used for other stellarator coil calculations.
The fixed field $\Bf$ is taken to be a purely toroidal field, approximating the field from a large number of toroidal field coils.
The mean field in the plasma region is 0.5 T, one third of the original NCSX design, as this is what can be supplied with the array of planar toroidal field coils built for NCSX.
The inner magnet surface is taken to be the NCSX vacuum vessel.
This vessel is not a uniform distance from the plasma boundary.
For this subsection we consider a uniform magnet thickness $d =$ 0.1 m. The weight $w$ in (\ref{eq:regularization}) is set to 1 until section \ref{sec:ports}.

\begin{figure}
  \centering
  \includegraphics[width=4in]{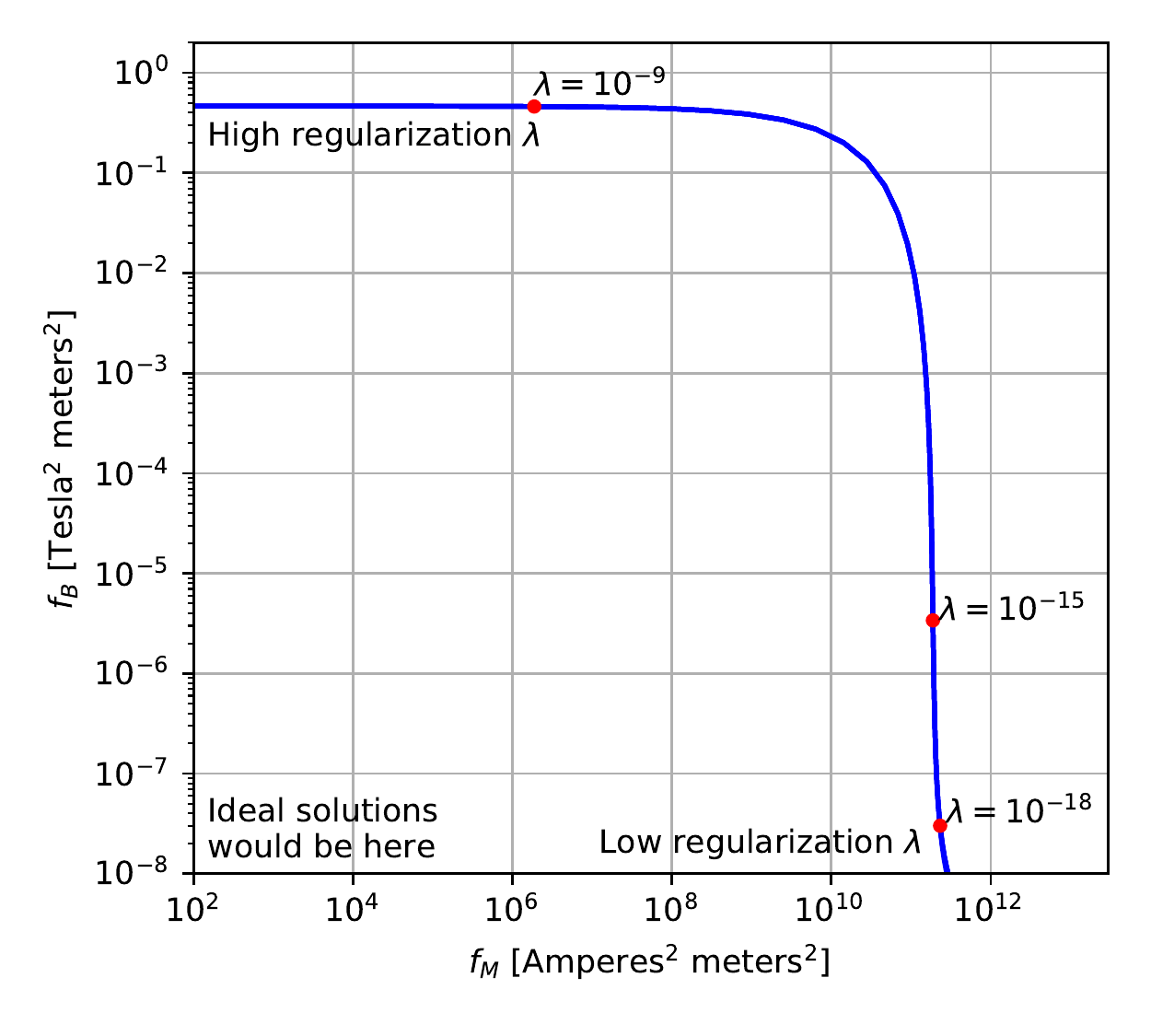}
  \caption{
Trade-off curve as the regularization parameter $\lambda$ is varied, at fixed magnet thickness $d = $0.1 m. Units of $\lambda$ are Tesla$^2$ / Ampere$^2$. Red dots show the solutions in figure \ref{fig:regularization}.
  }
\label{fig:Pareto}
\end{figure}

\begin{figure}
  \centering
  \includegraphics[width=6.1in]{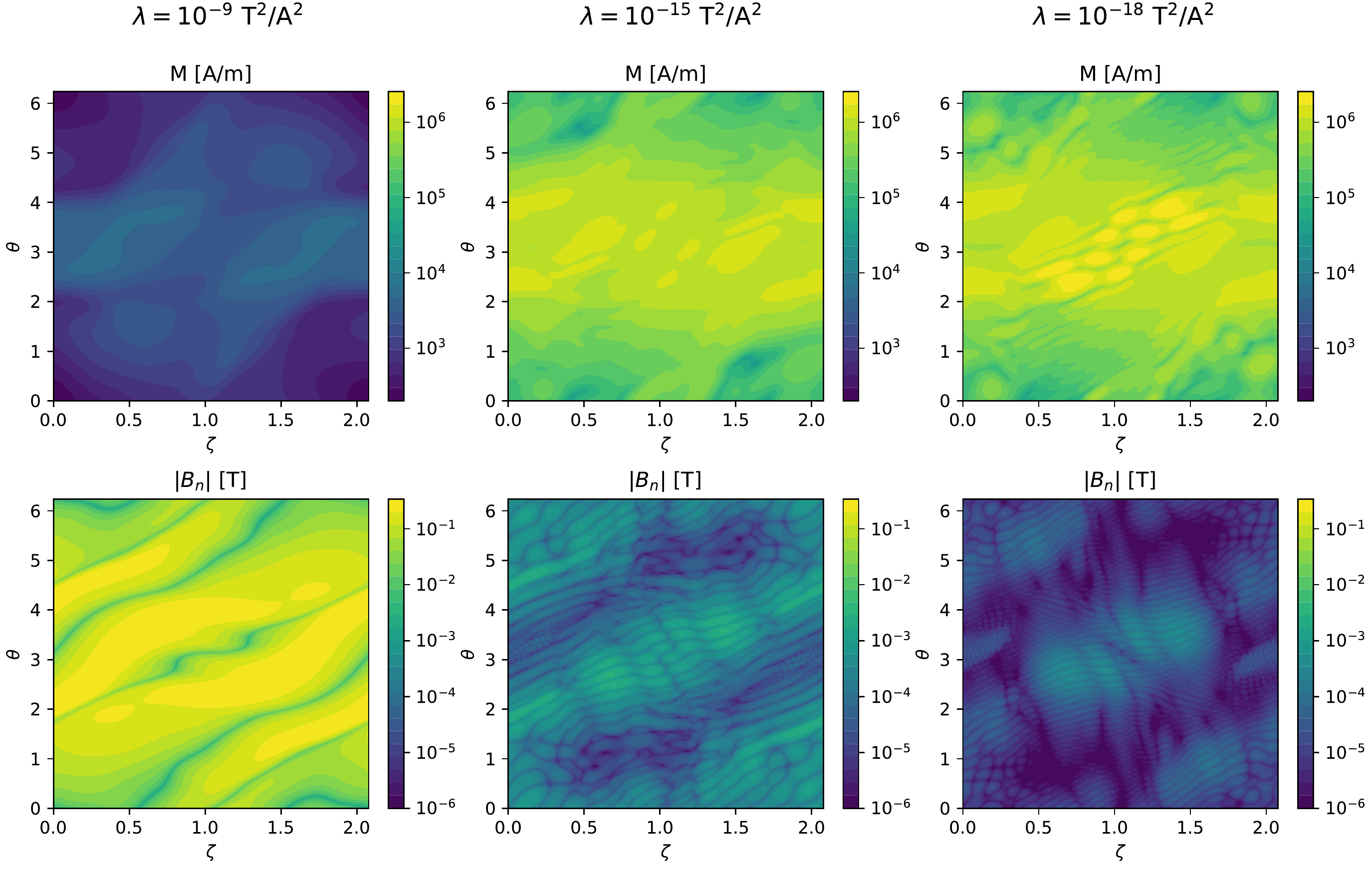}
  \caption{
Trends as the regularization parameter $\lambda$ is varied in the least-squares problem $\min f$ for fixed magnet thickness $d$. The three
solutions shown correspond to the red points in figure \ref{fig:Pareto}.
As regularization is reduced, the residual normal field $B_n = \left( \BPM + \Bf \right)\cdot \vect{n}$ on the target plasma surface is reduced, but at the expense of greater magnetization magnitude and finer structure in the magnetization.
  }
\label{fig:regularization}
\end{figure}

Figure \ref{fig:Pareto} shows the trade-off curve (``Pareto frontier'')
between $f_B$ and $f_M$ as the level of regularization $\lambda$ is varied.
Ideally both $f_B$ and $f_M$ would both be small, but a trade-off must be made: a small value of one of these quantities requires a large value for the other. 
The trade-off curve plotted is actually 5 curves overlaid, showing that factor-of-2 changes in each numerical resolution parameter has negligible effect on the solutions (Table \ref{tab:resolution}).
Three red points indicate solutions that are shown in detail in figure \ref{fig:regularization}

At large $\lambda$, the trade-off curve extends infinitely far to the left. In this limit, $f \approx \lambda f_M$, so the solution is $\vect{M} \to 0$. With no permanent magnets, $f_B$ has a nonzero value associated with the fixed field $\Bf$. 
At the other limit of small $\lambda$, arbitrarily small values of $f_B$ and arbitrarily large values of $f_M$ are obtained. (The curve eventually bends to the right but very large numerical resolution is required in this region, so only the converged section is displayed.) In this limit, the component of $\vect{B}$ normal to the target plasma surface is made arbitrarily small due to extremely large values of $M$. The regularization vanishes in this limit, so very short-scale patterns in $\vect{M}$ arise. 
A user must choose an intermediate value of $\lambda$ that balances magnet complexity against physics properties of the plasma configuration.

\begin{table}
\caption{\label{tab:resolution}
Resolution parameters for the five overlaid blue curves in figure \ref{fig:Pareto}. 
$N_{\theta}$ and $N_{\zeta}$: Number of grid points in the poloidal and toroidal angles on the plasma surface. 
$N_{\theta'}$ and $N_{\zeta'}$: Number of grid points in the poloidal and toroidal angles in the magnetization region. 
$m_{\max}$ and $n_{\max}$: Maximum Fourier mode numbers for the cylindrical components of $\vect{M}$.
$N_s$: Number of Gauss-Legendre points for integration over the radial coordinate $s$ in the magnetization region.
$n_{fp}$: Number of identical field periods.
}
\begin{indented}
\item[]\begin{tabular}{@{}ccccc}
\br
Run \# & $N_{\theta}=N_{\zeta}/n_{fp}$ & $N_{\theta'}=N_{\zeta'}/n_{fp}$ & $m_{\max}=n_{\max}$ & $N_s$ \\
\mr
1 & 128 & 128 & 32 & 3 \\
2 & 128 & 128 & 32 & 6 \\
3 & 256 & 128 & 32 & 3 \\
4 & 128 & 256 & 32 & 3 \\
5 & 128 & 129 & 64 & 3 \\
\br
\end{tabular}
\end{indented}
\end{table}

\subsection{Fixed-point iteration}
\label{sec:fixedPoint}

We now add the fixed-point iteration (\ref{eq:d_update}),
considering the same 0.5 T NCSX geometry from the previous subsection.
The iteration converges fastest when the number of degrees of freedom in $d$ is close to the number of degrees of freedom in each component of $\vect{M}$,
i.e. when $N_{\theta'} = 2 m_{\max}+1$ and $N_{\zeta'}/n_{fp}=2 n_{\max}+1$. 
Otherwise the spatial dependence of $d$ and $M$ in (\ref{eq:d_update})
does not match.
Therefore for this section we use the parameters of run 5 from table \ref{tab:resolution}. We  choose $\lambda = 10^{-15}$ T$^2/$A$^2$.
We also choose a target magnetization $M_t = 1.4$ T$/ \mu_0 \approx 1.114$ MA$/$m, achievable with rare-Earth magnets.

\begin{figure}
  \centering
  \includegraphics[width=6.1in]{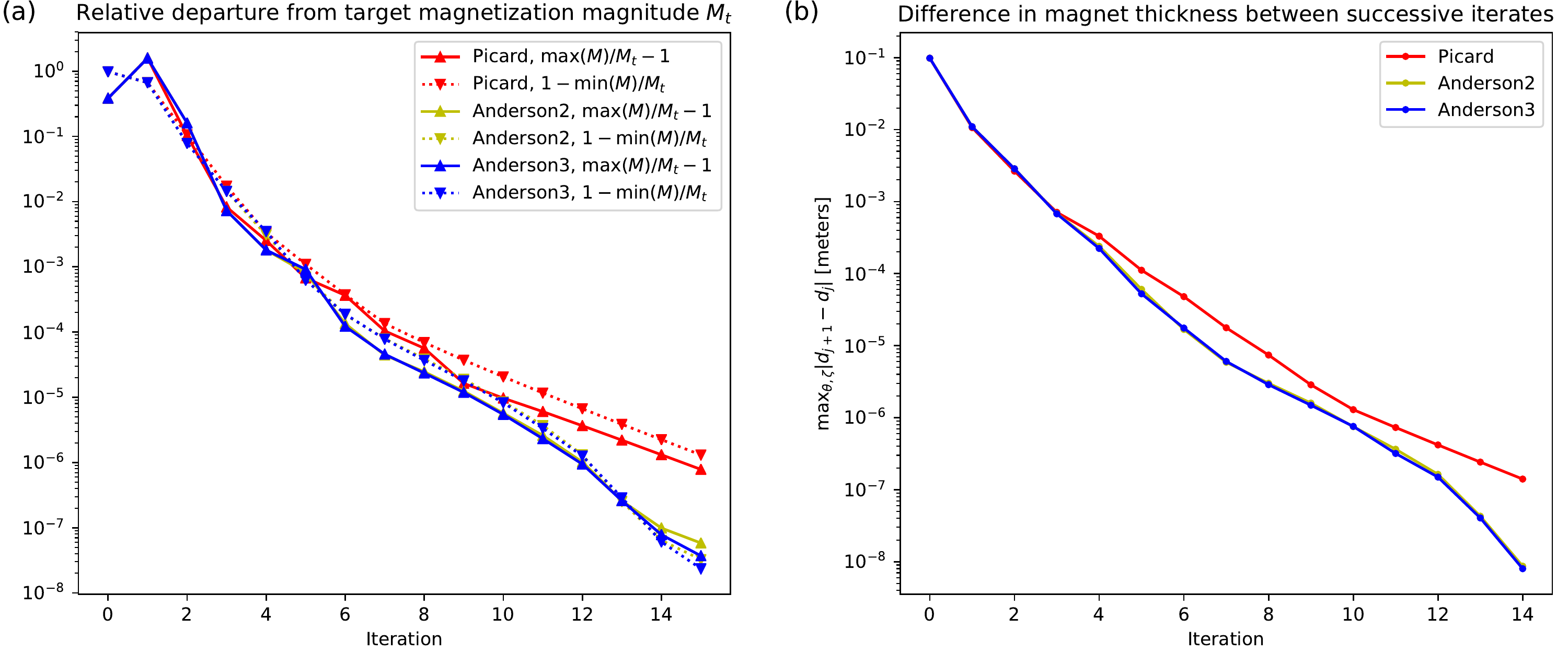}
  \caption{
Convergence of the fixed-point iterations, showing both the Picard
update (\ref{eq:d_update}) and its Anderson-accelerated variant.
The 2 or 3 after Anderson refers to the `depth' of Anderson acceleration.
(a) Both the maximum and minimum of $M$ over the magnetization volume converge to the desired value $M_t$, i.e. $M$ becomes uniform.
(b) The difference in the shape of the magnetization region between successive iterates converges to zero.
  }
\label{fig:d_convergence}
\end{figure}

Figure \ref{fig:d_convergence} shows the convergence of the fixed-point iterations. It can be seen that the minimum and maximum of $M$ over $V$ both quickly converge to the target $M_t$. Figure \ref{fig:d_convergence}.b shows the difference in $d$ between successive iterates, measured by the maximum over $\theta$ and $\zeta$ of $|d_{j-1} - d_j|$. The difference converges to zero, demonstrating that a fixed point has been found. 

Also shown in figure \ref{fig:d_convergence} are results when Anderson acceleration \cite{Anderson,Tonatiuh} is applied to the iteration. In Anderson acceleration, a linear combination of the previous few iterates is used instead of only the previous iterate. The extra computational cost of the Anderson step compared to (\ref{eq:d_update}) is so small as to be negligible, and here it provides a modest acceleration in convergence. 

\begin{figure}
  \centering
  \includegraphics[width=6.1in]{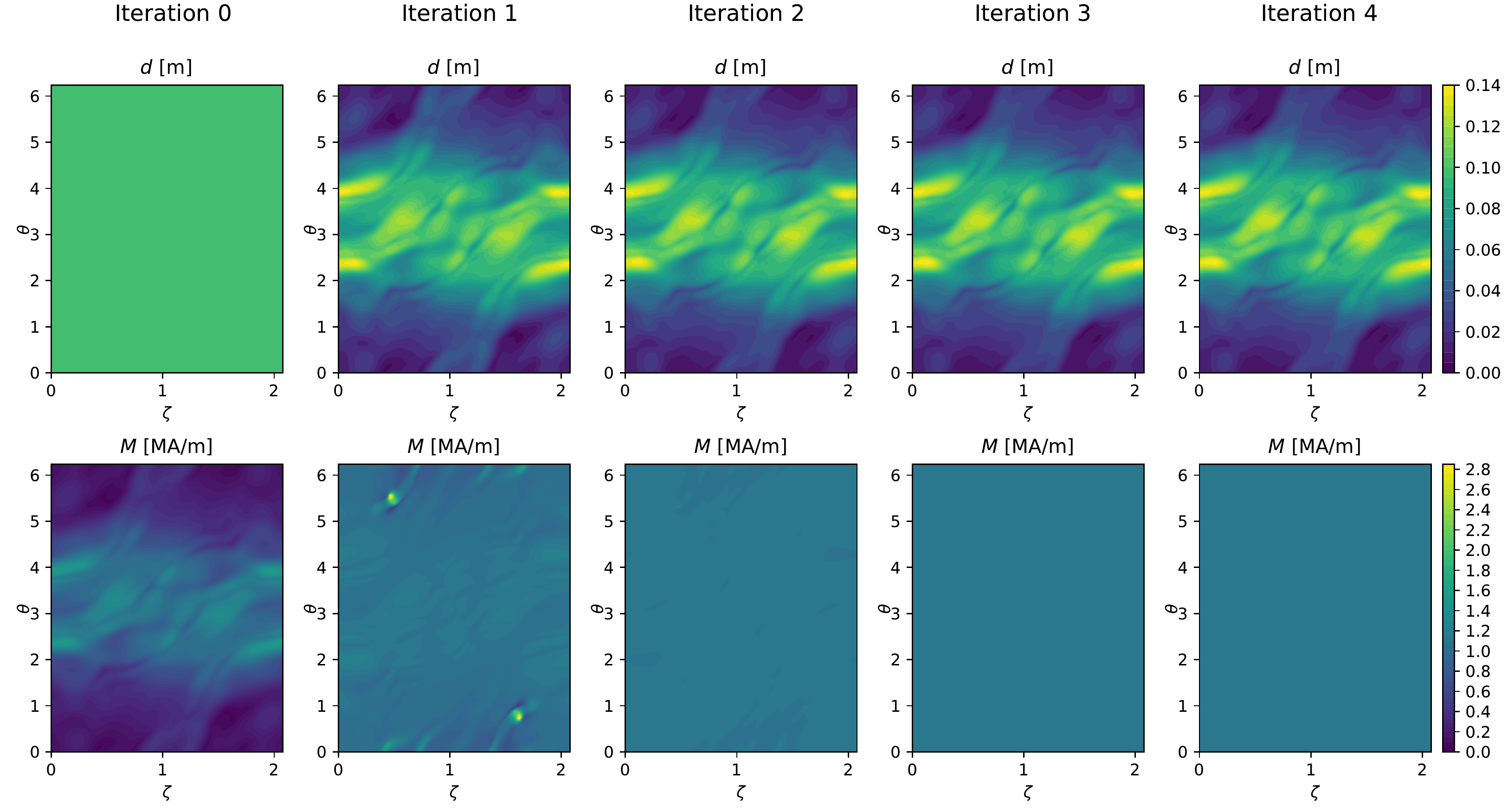}
  \caption{
Convergence of the magnet thickness $d$ and magnetization magnitude $M$ during the fixed-point iterations (\ref{eq:d_update}). By iteration 3, deviations of $M$ from the target value 1.1 MA/m are invisible on the scale of the figure.
  }
\label{fig:d_convergence_spatial}
\end{figure}

\begin{figure}
  \centering
  \includegraphics[width=6.1in]{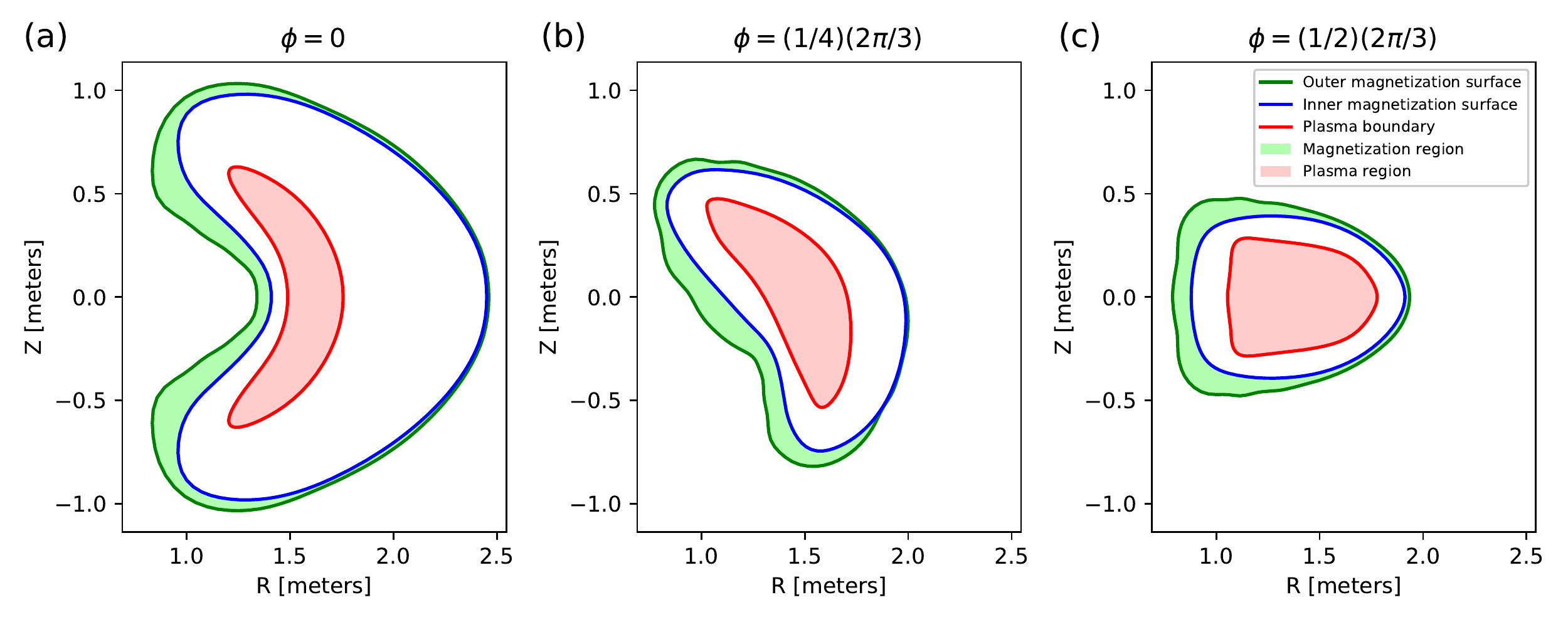}
  \caption{
Cross-sections of the REGCOIL\_PM solution for the NCSX example in section \ref{sec:fixedPoint}.
  }
\label{fig:XSecions_noPorts}
\end{figure}

The evolution of the spatial dependence of $d$ and $M$ is shown in figure
\ref{fig:d_convergence_spatial}. It can be seen that both $d$ and $M$ converge rapidly. 
By eye, $M$ is uniform and equal to $M_t$ by iteration 3, and changes to $d$ are hardly visible after iteration 1.
The final result for the shape of the magnet region is displayed in figure \ref{fig:XSecions_noPorts}.

\begin{figure}
  \centering
  \includegraphics[width=4in]{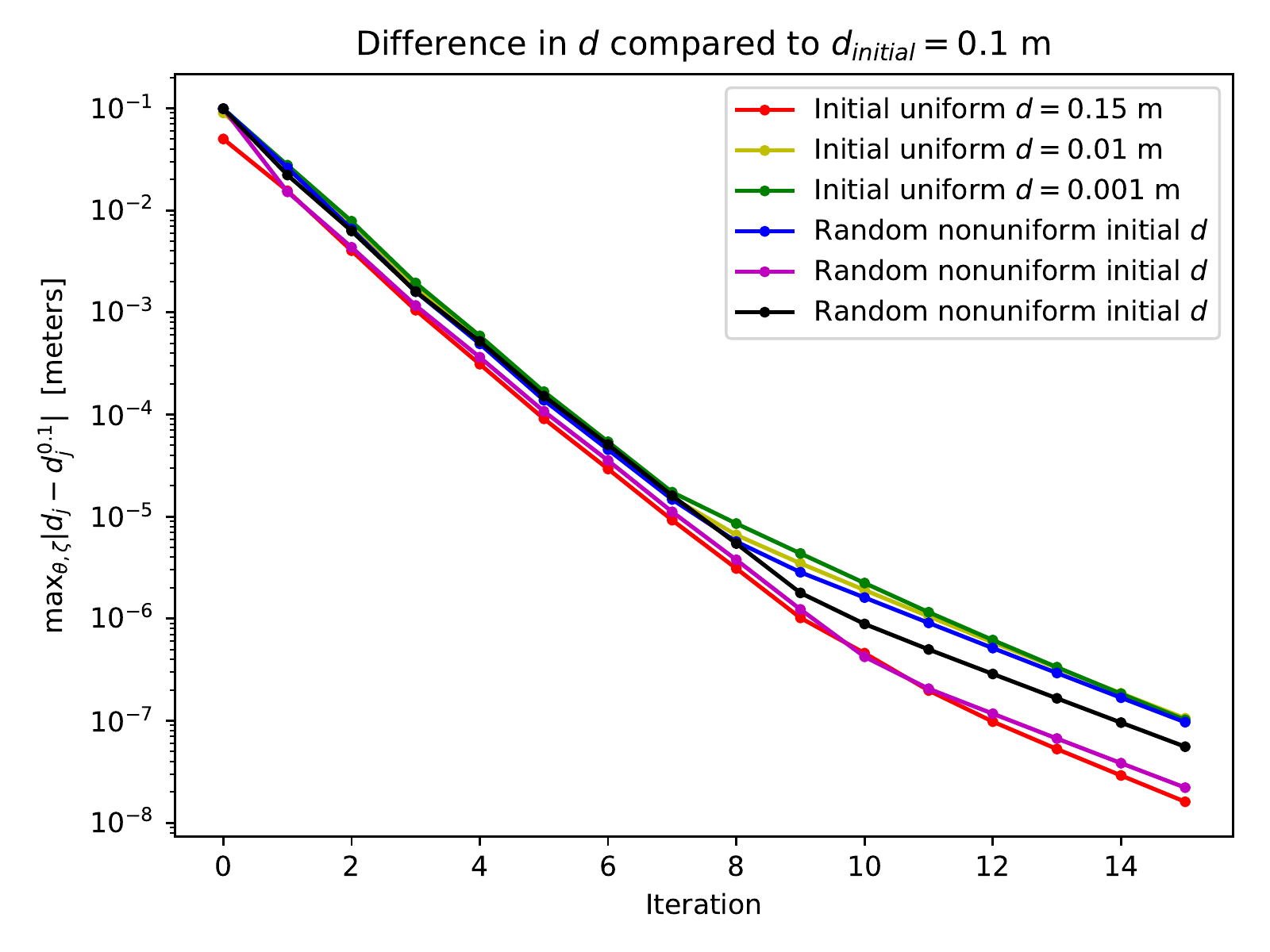}
  \caption{
For different choices of the initial magnet thickness $d$,
the iterations (\ref{eq:d_update}) converge to the same solution.
Here, $d_j^{0.1}$ denotes $d$ at iteration $j$ for a calculation initialized with a uniform $d=0.1$ m.
For all other initial conditions, $|d_j - d_j^{0.1}|$ is computed, and the maximum of this difference over $\theta$ and $\zeta$ is plotted.
 The difference converges towards zero. }
\label{fig:compareDifferentInitialD}
\end{figure}

It is not obvious that for any choice of initial $d$, the iteration is stable and the fixed point obtained is the same. 
However it appears that a unique solution exists in practice, at least for the examples in this paper.
The calculation of this section was repeated with various uniform initial $d \in \{0.001, 0.01, 0.15\}$ meters. (Larger values are not permitted because the outer surface begins to self-intersect as $\sqrt{g}$ crosses through zero.) 
The calculation was also repeated taking the initial $d$ to have a random variation in $\theta$ and $\zeta$ within $[0, 0.1]$ m.
As shown in figure \ref{fig:compareDifferentInitialD}, differences between these differently-initialized calculations converged steadily towards zero as the iterations proceeded. 
This behavior is in contrast to the formulations in \cite{ZhuTopology,Hammond} in which dependence on the initial condition was observed.
The independence of REGCOIL\_PM results from the initial condition is advantageous, since a user need not worry about how best to select the initial condition.

\section{Verification for Halbach cylinders}
\label{sec:benchmark}


A satisfying property of the mathematical
formulation of section \ref{sec:formulation}
is that it is consistent with the analytic
solution for cylindrical multipole magnets (``Halbach cylinders'')
described by Halbach \cite{Halbach}. A comparison
with this analytic result also serves as a useful
test of the numerical implementation of section \ref{sec:numerical}.


\subsection{Analytic solution}
\label{sec:benchmarkAnalytic}

\begin{figure}
  \centering
  \includegraphics[width=5in]{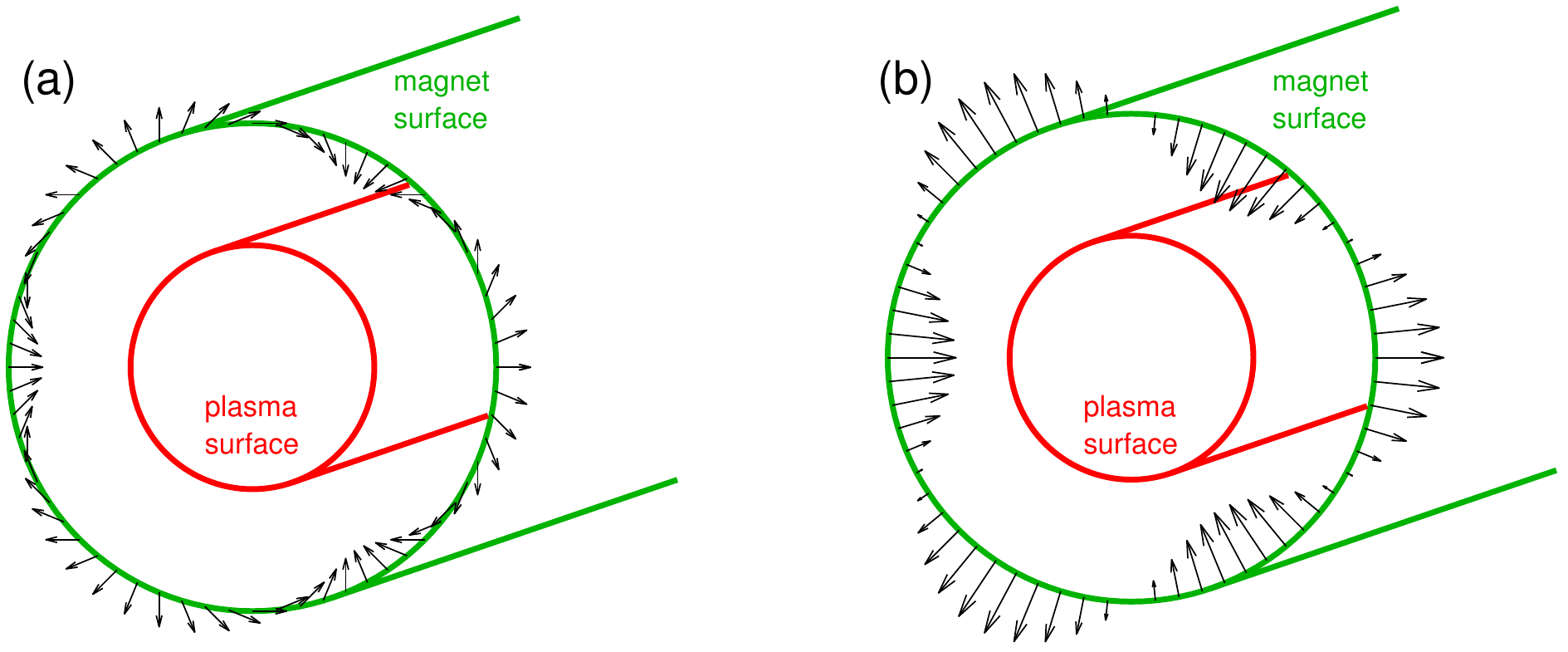}
  \caption{
The two configurations in cylindrical geometry analyzed in section \ref{sec:benchmarkAnalytic}. (a) 
Dipoles with uniform magnitude but varying direction, yielding the Halbach cylinder solution.
(b) Dipoles oriented normal to the magnet surface with varying magnitude.
  }
\label{fig:Halbach}
\end{figure}

We first derive the analytic solution by a different method
than in \cite{Halbach} to highlight the parallels with stellarator
magnet optimization. We consider two concentric infinite cylinders,
an inner one with radius $a$ analogous to the plasma surface,
and an outer one with radius $b>a$ analogous to a thin magnet volume.
This configuration can be imagined as a high-aspect-ratio
limit of an axisymmetric system, so the angle around the cylinder $\theta$
is a poloidal angle.
Let us try to arrange magnetic dipoles on the outer surface 
in order to create a normal magnetic field
\begin{align}
\label{eq:Bn_benchmark}
    B_n = \vect{B}\cdot\vect{n} = \bar{B} \cos(\ell \theta)
\end{align}
on the plasma surface, where $\ell$ is a given integer.
In other words, suppose there is a fixed normal field $B_{n,f}=-\bar{B} \cos(\ell \theta)$, and we wish to introduce dipoles to obtain $f_B=0$.
We will consider two possible arrangements of dipoles, shown in figure \ref{fig:Halbach}:
first, dipoles of uniform magnitude but arbitrary direction,
as in a REGCOIL\_PM solution; and second, dipoles oriented normal to the outer
surface but with arbitrary magnitude.
This second case is considered because dipoles oriented normal to a surface have been considered in recent papers \cite{ZhuPerpendicular,Hammond}.
We will show the magnetization magnitude in the first approach is half of the maximum magnetization in the second.

Outside of the region of dipoles, the magnetic field can be written
$\vect{B} = \nabla\Phi$ for a scalar potential $\Phi$. The potential
for a single point dipole is $\Phi = \Phi_d$ where
\begin{align}
\label{eq:dipolePotential}
    \Phi_d(\vect{r})=-\frac{\mu_0 (\vect{r}-\vect{r}')\cdot\vect{m}}{4\pi |\vect{r}-\vect{r}'|^3}.
\end{align}
Here, $\vect{m}$ is the magnetic moment, $\vect{r}'$ is the position vector of the dipole, and $\vect{r}$ is the observation location. The gradient of (\ref{eq:dipolePotential}) gives the expected field
\begin{align}
    \vect{B}(\vect{r}) = 
    \frac{\mu_0}{4\pi |\vect{r} - \vect{r}'|^3} \left[ \frac{3(\vect{r}- \vect{r}')(\vect{r}- \vect{r}')\cdot\vect{m}}{|\vect{r} - \vect{r}'|^2}
    -\vect{m}\right].
\end{align}
We introduce cylindrical coordinates $(r,\theta,z)$ and Cartesian coordinates $(x,y,z)$ with the $z$ axis along the axis of the cylinder,
and associated unit vectors $(\vect{e}_x, \; \vect{e}_y, \; \vect{e}_z)$. Again we use primes to indicate coordinates on the magnet surface, so the evaluation and source positions are
\begin{align}
    \vect{r}&=r\cos\theta \vect{e}_x + r \sin\theta \vect{e}_y+z\vect{e}_z,\\
    \vect{r}'&=b\cos\theta' \vect{e}_x + b \sin\theta' \vect{e}_y+z'\vect{e}_z. \nonumber
\end{align}
Supposing the dipoles cover the surface with a uniform number density $\eta$ (units of 1$/$area), then the total potential is 
\begin{align}
\label{eq:generalPotential}
    \Phi(\vect{r})=\eta \int d^2r' \Phi_d
=    
    -\frac{\mu_0 \eta b}{4\pi } \int_0^{2\pi} d\theta' \int_{-\infty}^{\infty}dz'
    \frac{(\vect{r}-\vect{r}')\cdot\vect{m}(\theta')}{|\vect{r}-\vect{r}'|^3}.
\end{align}

For the first of the two configurations, shown in figure \ref{fig:Halbach}.a, we consider dipoles
\begin{align}
\label{eq:uniformMagnitudeDipoles}
    \vect{m} = \hat{m} \cos(n\theta') \vect{e}_x + \hat{m}\sin(n\theta') \vect{e}_y,
\end{align}
for some integer $n$ and constant $\hat{m}$ so $|\vect{m}|$ is uniform. The integral (\ref{eq:generalPotential}) for this case is evaluated in  \ref{apx:integrals}, with the result
\begin{align}
\label{eq:potentialUniform}
    \Phi_{uni} = \mu_0 \eta \hat{m} \left(\frac{r}{b}\right)^{n-1} \cos((n-1)\theta)
\end{align}
for $r<b$. We will not need the field for $r>b$. The magnetic field normal to the inner surface is then
\begin{align}
    B_n=\left(\frac{\partial\Phi}{\partial r}\right)_{r=a} = \frac{\mu_0 \eta \hat{m}(n-1)}{b} \left( \frac{a}{b}\right)^{n-2} \cos((n-1)\theta).
    \label{eq:uniformMagnitude_thin}
\end{align}
Comparing this result to (\ref{eq:Bn_benchmark}), we see the desired field is produced on the plasma surface if we choose $n=\ell+1$ and
\begin{align}
\label{eq:uniformDipoleResult}
    \hat{m}=\frac{\bar{B} b}{\mu_0 \eta \ell} \left(\frac{b}{a}\right)^{\ell-1}.
\end{align}

Expression (\ref{eq:uniformMagnitude_thin}) for a thin layer of dipoles can be extended to a formula for a finite-thickness magnet with inner radius $b_1$ and outer radius $b_2$ by writing $\eta \hat{m} = M \, db$ and integrating in $b$ over $[b_1, \, b_2]$. The result is
\begin{align}
    B_n= \frac{\mu_0 M (n-1)}{(n-2)} \left( \frac{a}{b_1}\right)^{n-2}
    \left[1-\left(\frac{b_1}{b_2}\right)^{n-2}\right]\cos((n-1)\theta).
\end{align}
This result is equivalent to the radial component of (21a) in \cite{Halbach}, Halbach's multipole, noting the following substitutions: $N \to n-1$, $B_r \to \mu_0 M$, $r_{1,2} \to b_{1,2}$, and $\varphi \to \theta$. Equivalently, the magnetization required to produce the field (\ref{eq:Bn_benchmark}) is
\begin{align}
\label{eq:HalbachM}
M=\frac{\bar{B}(\ell-1)}{\mu_0 \ell} \left(\frac{b_1}{a}\right)^{\ell-1} \left[1-\left(\frac{b_1}{b_2}\right)^{\ell-1}\right]^{-1}.    
\end{align}

Since this dipole configuration produces the desired $B_n$ (\ref{eq:Bn_benchmark}) exactly, 
then when it is added to the aforementioned equal and opposite fixed field $B_{n,f}=-\bar{B}\cos\ell\theta$ one obtains $f_B=0$.
Therefore this dipole configuration is a solution of the RECGOIL\_PM least-squares step in the limit of small $\lambda$. Furthermore, 
since $M$ is uniform, this configuration is a fixed point of the Picard iteration. Therefore this configuration is a fixed point of the overall REGCOIL\_PM algorithm.

We can compare this first configuration of dipoles with
the second configuration, in which the dipole directions
are constrained to lie in the direction normal to the surfaces,
now allowing $M$ to vary with $\theta'$. This second configuration
is illustrated in figure \ref{fig:Halbach}.b. We assume the magnitude
of the dipoles is $|\vect{m}|=\bar{m} \cos\ell\theta'$ for
some constant $\bar{m}$, so
\begin{align}
\label{eq:normalDipoles}
    \vect{m}=\bar{m}\cos(\ell\theta') \left( \vect{e}_x\cos\theta' + \vect{e}_y\sin\theta'\right).
\end{align}
This expression is substituted into (\ref{eq:generalPotential}),
and after evaluating the integrals as shown in  \ref{apx:integrals}, one finds
$\Phi=\Phi_{nor}$ for
\begin{align}
\label{eq:potentialNormal}
    \Phi_{nor}=\frac{\mu_0 \eta \bar{m}}{2}\cos(\ell\theta) \left(\frac{r}{b}\right)^{\ell}
\end{align}
for $r<b$. We will not need the field for $r>b$. The field normal to the plasma surface is then
\begin{align}
    B_n = \left(\frac{\partial\Phi_{nor}}{\partial r}\right)_{r=a}
    =\frac{\mu_0 \eta \bar{m} \ell}{2 b} \cos\ell\theta \left(\frac{a}{b}\right)^{\ell-1}.
\end{align}
Comparing this expression to (\ref{eq:Bn_benchmark}) it can be seen
that the desired field on the plasma surface is produced if the maximum
dipole magnitude is
\begin{align}
    \bar{m} = \frac{2 \bar{B} b}{\mu_0 \eta \ell} \left(\frac{b}{a}\right)^{\ell-1}.
    \label{eq:normalDipolesResult}
\end{align}
The fact that the dipole arrangements (\ref{eq:uniformMagnitudeDipoles})
and (\ref{eq:normalDipoles}) can both produce the same field (\ref{eq:Bn_benchmark}) on the plasma reflects the significant freedom available in choosing the magnets for a given
stellarator.
Comparing (\ref{eq:normalDipolesResult}) to (\ref{eq:uniformDipoleResult}), we see that the required maximum dipole
magnitude is twice as large when the dipoles are constrained
to lie normal to the magnet surface compared to the arbitrary-orientation case.


\subsection{Numerical solution}
\label{sec:benchmarkNumerical}

We now compare the analytic result (\ref{eq:HalbachM}) to numerical
calculations with REGCOIL\_PM. Since the code is written for
toroidal geometry rather than cylindrical geometry, we choose
a very large but finite aspect ratio, with major radius 30 m, $b_1=1$ m, $a=1/3$ m. We also choose $\bar{B}=1$ T and, initially, $\ell=3$. For this section we neglect the Picard iteration to focus on the 
behavior of the regularized least-squares problem, fixing $d=1$ mm. We use the following resolution parameters: 96 grid points poloidally, 512 grid points toroidally, 24 Fourier modes poloidally, and 2 grid points radially.

\begin{figure}
  \centering
  \includegraphics[width=6.1in]{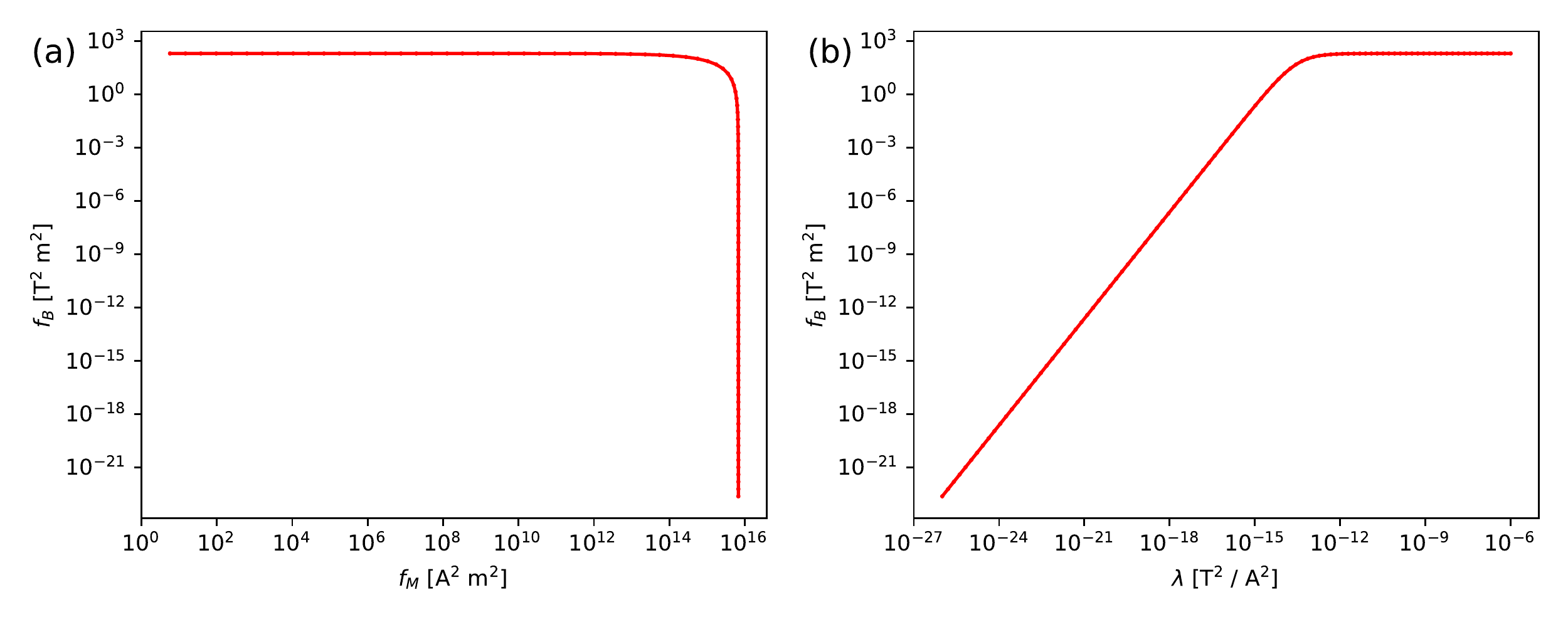}
  \caption{
Scan of the regularization parameter $\lambda$ for the Halbach cylinder benchmark problem of section \ref{sec:benchmarkNumerical}. (a) Pareto trade-off curve. (b) Normal magnetic field error vs $\lambda$.
  }
\label{fig:benchmark_Pareto}
\end{figure}

First, the behavior of the regularized least-squares solution is examined as the regularization parameter $\lambda$ is varied. As shown in figure (\ref{fig:benchmark_Pareto}), as $\lambda$ is decreased below $10^{-14}$ T$^2$/A$^2$, 
the normal field error $f_B$ can be made
arbitrarily small, indicating the permanent magnets exactly produce the desired field.
This regime corresponds to the vertical part of the Pareto curve in figure \ref{fig:benchmark_Pareto}.a.
For $\lambda$ above this threshold value, 
the problem becomes over-regularized, with the regularization term forcing the magnetization to be very small such that the dipoles do not significantly cancel the fixed field. This regime corresponds to the horizontal part of the Pareto curve in 
figure \ref{fig:benchmark_Pareto}.a.
For the rest of this section we focus on values of $\lambda$ below the threshold,
for which the magnet distribution and $f_M$
are insensitive to $\lambda$, and $f_B$ is very small.

\begin{figure}
  \centering
  \includegraphics[width=2.5in]{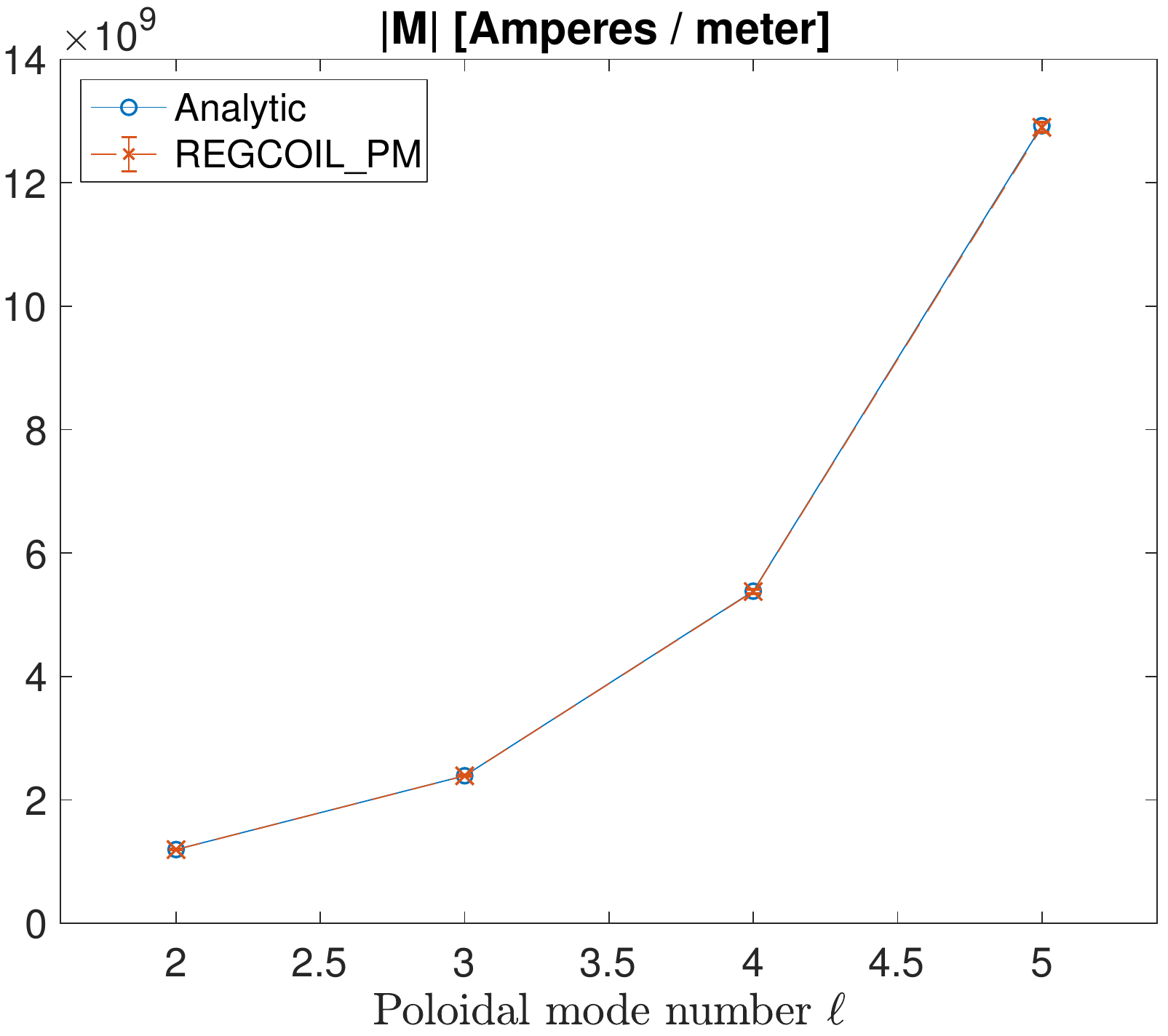}
  \hspace{0.2in}
  \includegraphics[width=2.5in]{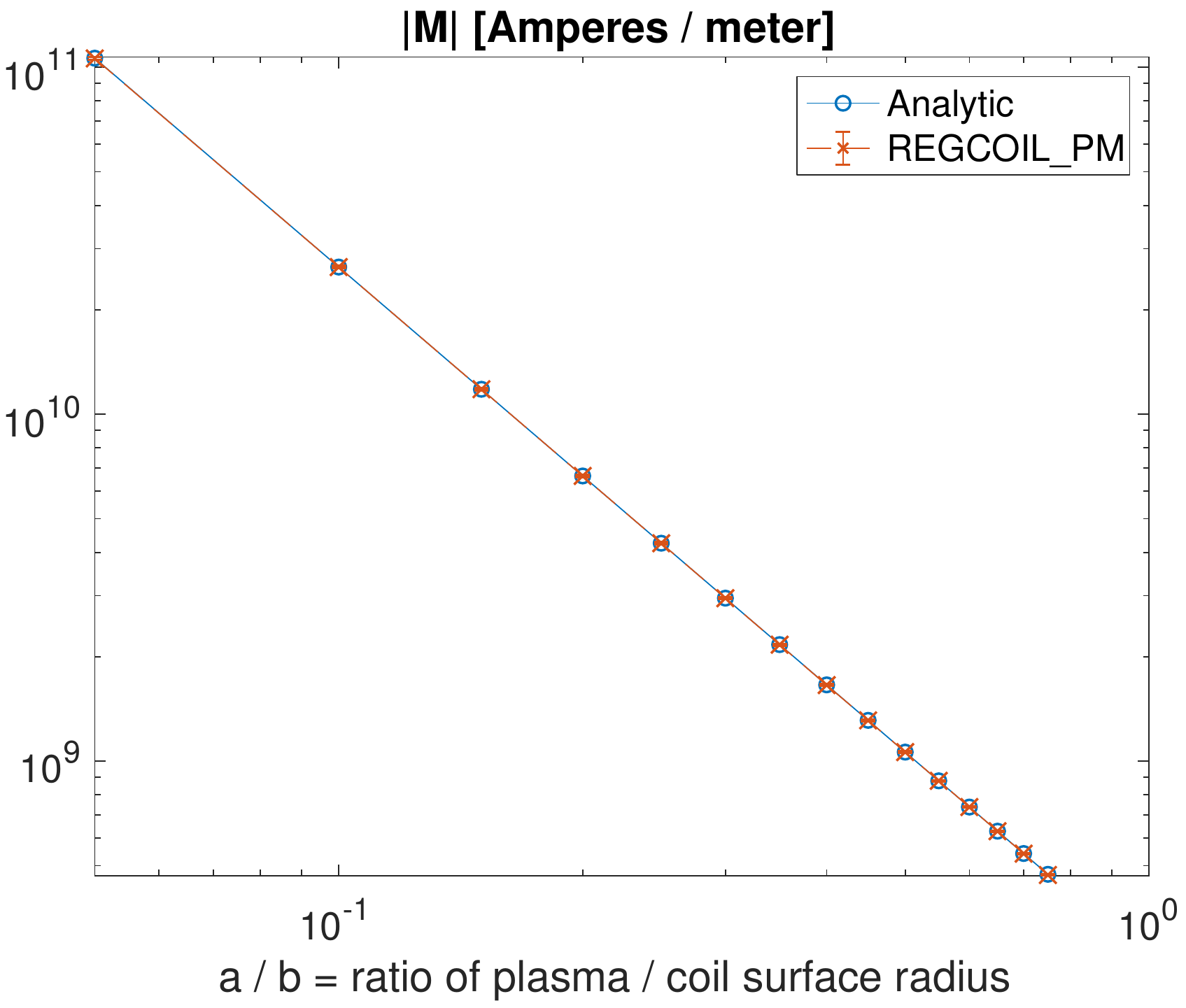}
  \caption{
Comparison of the analytic result (\ref{eq:HalbachM}) for a Halbach cylinder to  REGCOIL\_PM numerical calculations, as described in section \ref{sec:benchmarkNumerical}.
  }
\label{fig:benchmark_scans}
\end{figure}

Next, figures \ref{fig:benchmark_scans}.a-b show a comparison of the magnetization computed by REGCOIL\_PM to the analytic result 
(\ref{eq:HalbachM}), as $\ell$ or $a$ are varied. In both figures, error bars are given for the numerical results, displaying $\pm 1$ standard deviation of $M$ as $\theta'$ is varied over $[0,\,2\pi]$ and $\lambda$ is varied over $10^{-26} - 10^{-18}$ T$^2$/A$^2$. The error bars
are barely visible, indicating that $M$ is found to be uniform and independent of the regularization, as it should be. Extremely close agreement is found between the analytic and numerical results.

\begin{figure}
  \centering
  \includegraphics[width=3.5in]{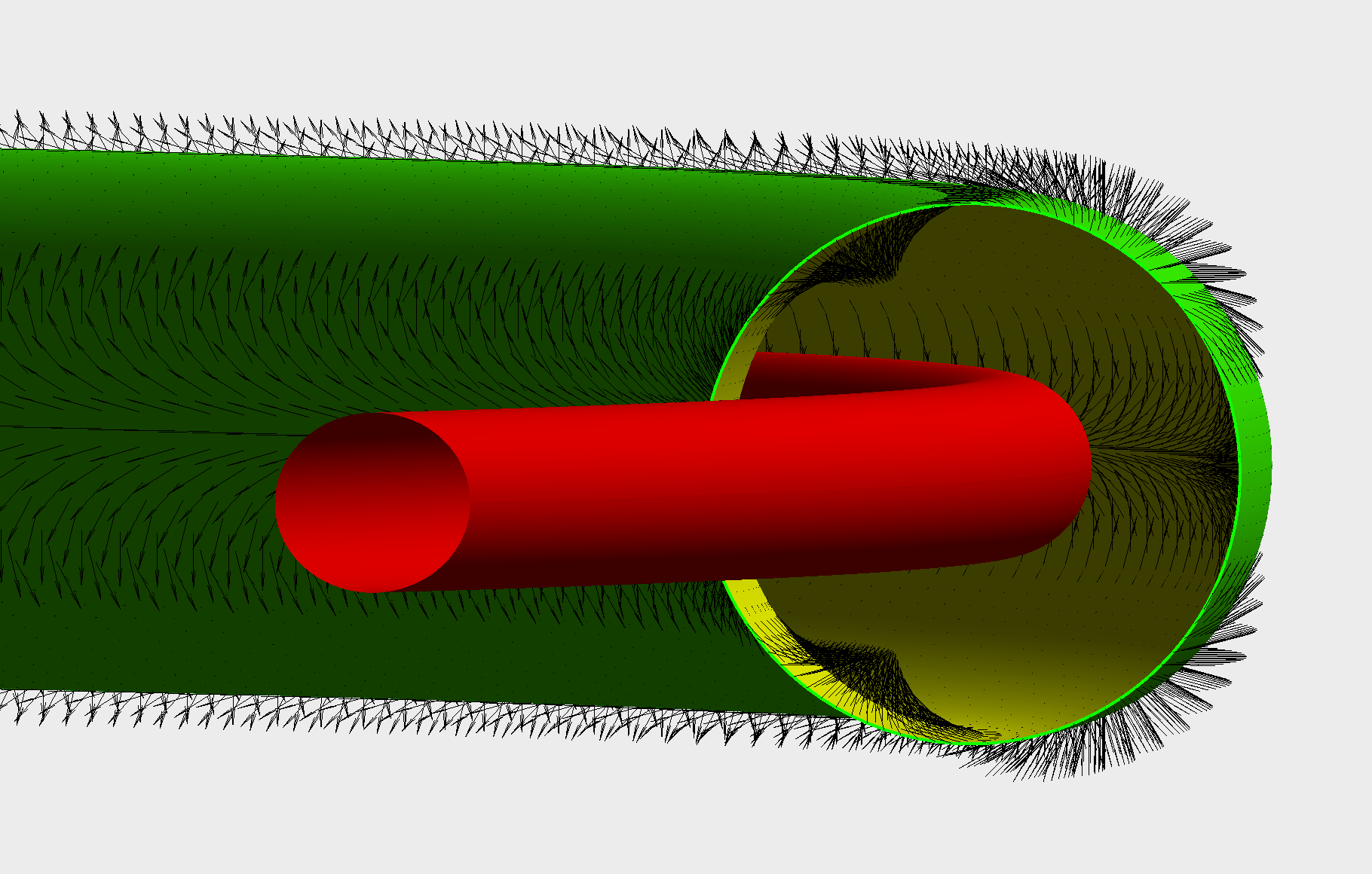}
  \caption{
The REGCOIL\_PM procedure can reproduce the Halbach cylinder solution, as described in section \ref{sec:benchmarkNumerical}.
Black arrows indicate the magnetization.
Here, $\ell=3$, the plasma surface is shown in red, and the magnet region is shown in green.
  }
\label{fig:benchmark3D}
\end{figure}

Finally, figure \ref{fig:benchmark3D}
displays a 3D rendering of the numerical
solution for $\ell=3$, $a=1/3$ m.
The magnetization vector is displayed with black arrows. It can be seen that the REGCOIL\_PM procedure has indeed ``discovered'' the
Halbach solution of figure \ref{fig:Halbach}.a.

\section{NCSX example}
\label{sec:NCSX}

We now further develop and analyze the NCSX example. In the following subsections, we demonstrate the ability to remove magnet in regions to make room for ports, free-boundary equilibria using the permanent magnets, and the difficult in raising the field magnitude.
Finally we present a comparison to the different algorithm of ref \cite{ZhuTopology}.


\subsection{Ports}
\label{sec:ports}

It is infeasible to surround the plasma completely with permanent magnets, since access is required for heating, diagnostics, and maintenance. We therefore now show how regions of the permanent magnets can be removed for ports. The feasibility of including ports in the 0.5 T NCSX configuration was examined previously in \cite{ZhuTopology, ZhuPerpendicular, Hammond}.

\begin{figure}
  \centering
  \includegraphics[width=4in]{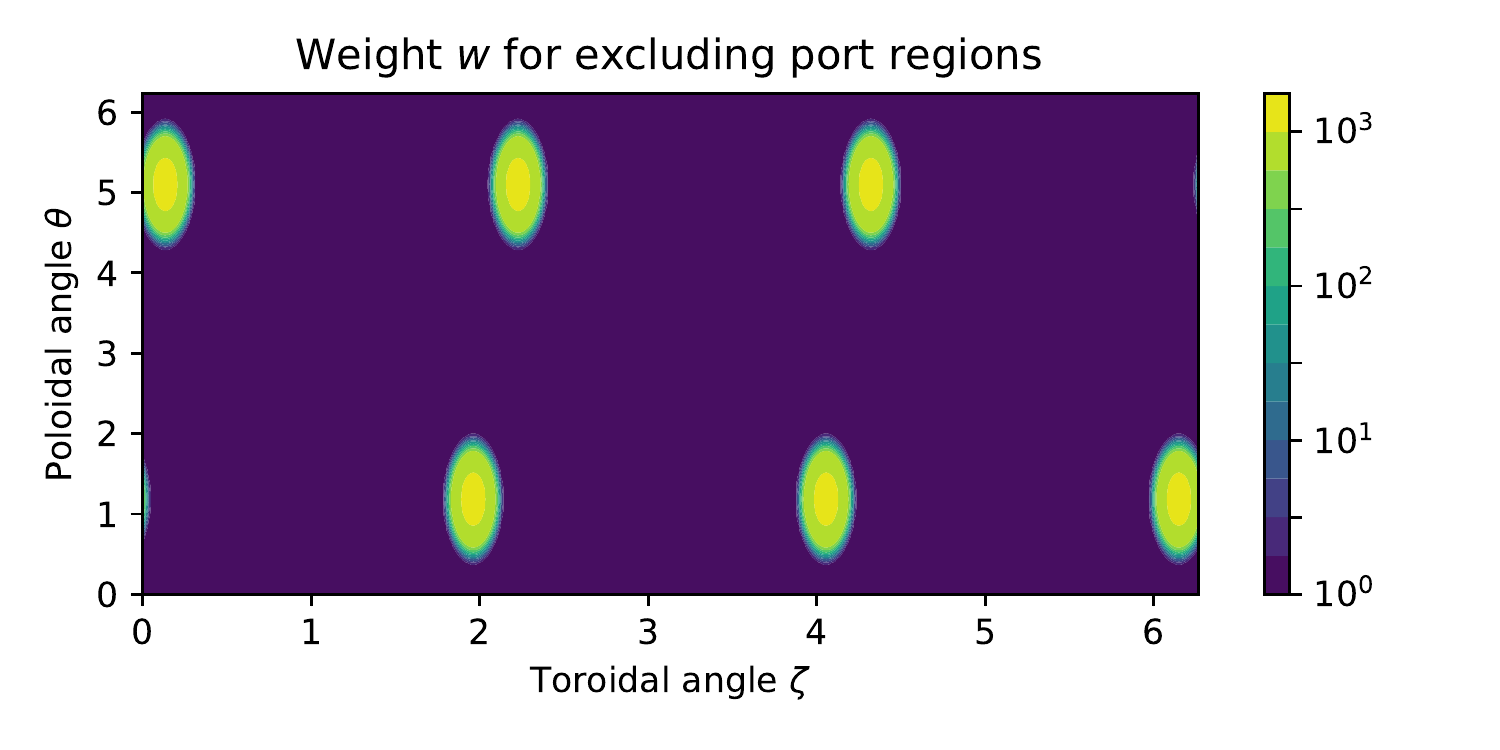}
  \caption{
The weight function $w(\theta, \, \zeta)$ used to exclude permanent magnets at the locations of ports for the NCSX example.
  }
\label{fig:ports_weight}
\end{figure}

Port regions are selected in REGCOIL\_PM by increasing the local value of the weight $w$ in (\ref{eq:regularization}).
For the example here we choose the following function for the weight:
\begin{align}
    w(\theta, \, \zeta) = &1 +  \sum_j \frac{A_j}{2}\left[ 1 + \tanh \left(s_j \left[ 1 - \frac{2}{\Delta\theta_j^2} \left[ 1-\cos(\theta-\theta_{0,j}) \right] 
    \right. \right. \right. \\
  & \hspace{2in}  \left. \left. \left.
    - \frac{2}{n_{fp}^2 \Delta\zeta_j^2} \left[ 1-\cos(n_{fp}\zeta-n_{fp}\zeta_{0,j}) \right] 
    \right] \right) \right] 
    \nonumber
\end{align}
This function is appropriately periodic in the two angles,
and the sum over $j$ allows multiple ports to be included.
Port $j$ is centered at $\theta = \theta_{0,j}$ and $\zeta = \zeta_{0,j}$, while the extent of the ports in $\theta$ and $\zeta$ is controlled by
$\Delta \theta_j$ and $\Delta \zeta_j$. The parameter $s_j$ controls
the sharpness of the transition from $w \approx 1$ to $w \gg 1$.
In the example here, we choose port 1 to have $\theta_{0,1}=0.4$, $\zeta_{0,1}=1.7$, $\Delta\theta_1 = 0.4$, $\Delta\zeta_1=0.2$, $A_1=1000$, and $s_1=5$. 
Additional ports are included with the same parameters but at stellator-symmetric and $n_{fp}$-symmetric locations. 
These values are chosen to align the ports with the regions of lowest magnet thickness, which are at the outboard side.
The resulting $w$ function is shown in figure \ref{fig:ports_weight}.

\begin{figure}
  \centering
 \includegraphics[width=6.1in]{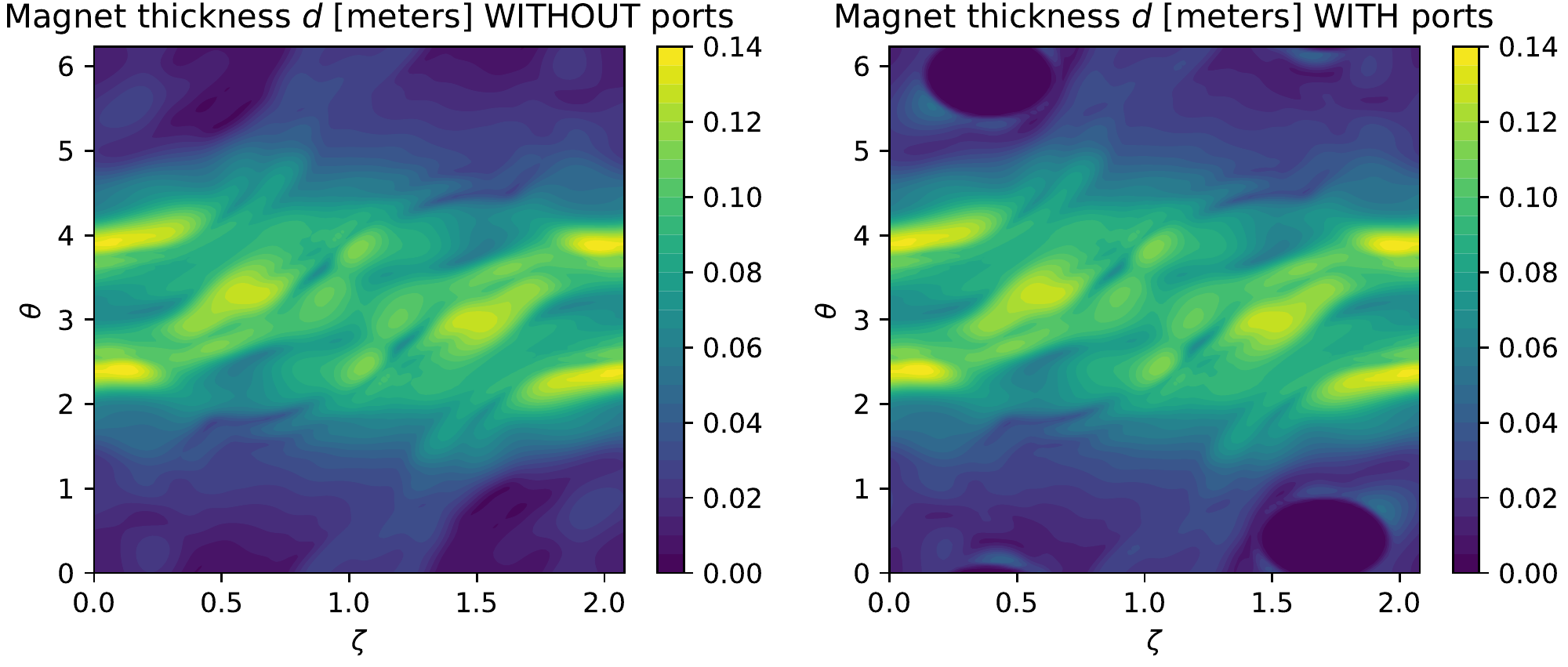}
  \caption{
Comparison of the magnet thickness computed by REGCOIL\_PM for the NCSX example without and with ports.
  }
\label{fig:ports_noports_comparison}
\end{figure}

\begin{figure}
  \centering
  \includegraphics[width=6.1in]{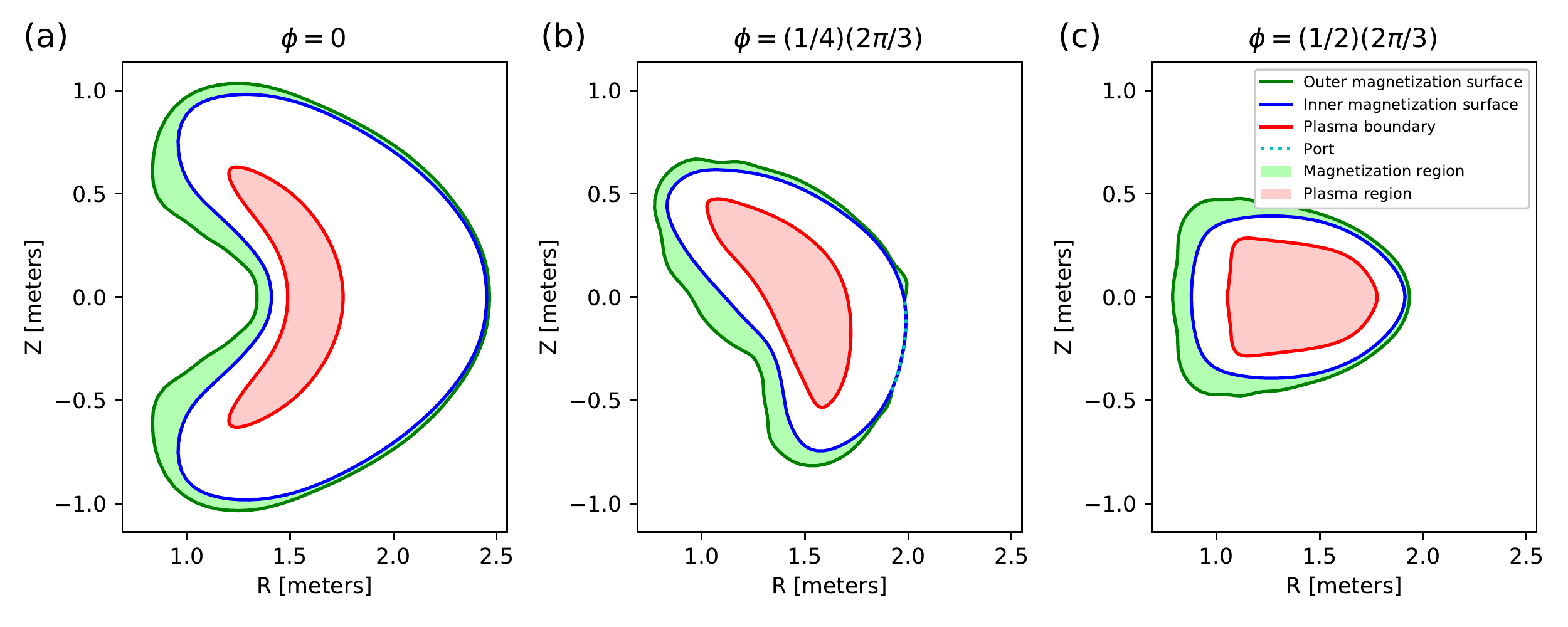}
  \caption{
REGCOIL\_PM solution for the NCSX example with ports.
  }
\label{fig:xsections}
\end{figure}

The REGCOIL\_PM solution with ports is displayed in figures \ref{fig:ports_noports_comparison} and \ref{fig:xsections}. The same regularization parameter is used as in section \ref{sec:fixedPoint}, $\lambda=10^{-15}$ T$^2/$A$^2$. It can be seen that the change to the magnet geometry is minor. A slight thickening of the magnet volume around the edge of the port is apparent. When ports are included, the volume of permanent magnets increases only slightly, from 2.012 m$^3$ to 2.025 m$^3$. The maximum $B_n$ also increases only slightly, from 0.00299 T without ports to 0.00303 T with ports. 
These results indicate it is likely that ports can be included in permanent magnet stellarators, at least in some locations.

\begin{figure}
  \centering
  \includegraphics[width=5.5in]{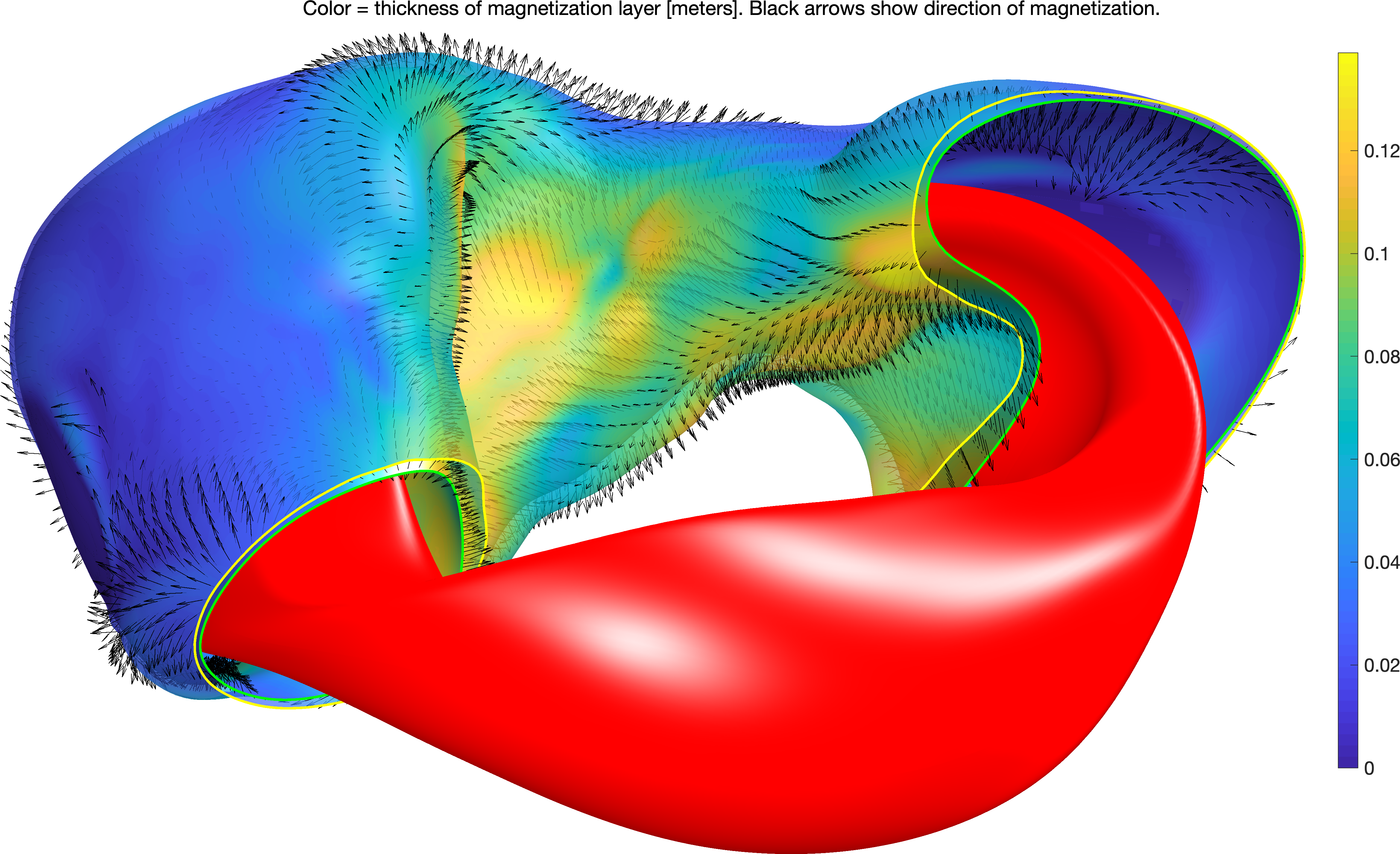}
    \includegraphics[width=5.5in]{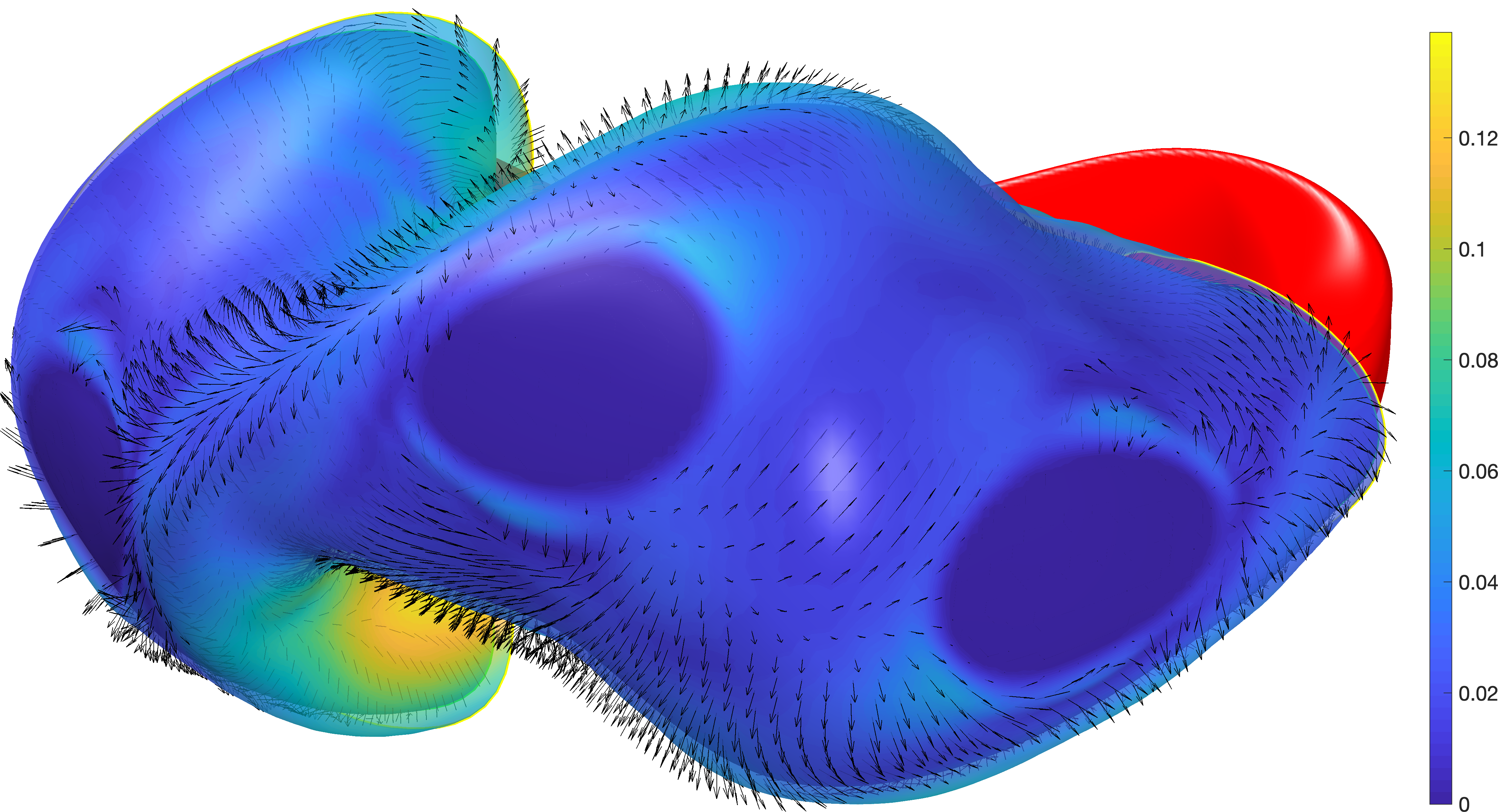}
  \caption{
The REGCOIL\_PM solution for the NCSX example with ports, viewed from two angles. The red surface is the plasma boundary. The inner and outer magnet boundaries $I$ and $O$ are shown, with their local color indicating $d$. Everywhere in the magnet region except for the ports, arrows of uniform length display the direction of $\vect{M}$.
  }
\label{fig:3D}
\end{figure}

A three-dimensional rendering of the REGCOIL\_PM solution with ports is shown in figure \ref{fig:3D}. In the magnetization region, arrows with uniform length are drawn everywhere except the ports to show the direction of $\vect{M}$.

\subsection{Free-boundary equilibria}

\begin{figure}
  \centering
  \includegraphics[width=6.1in]{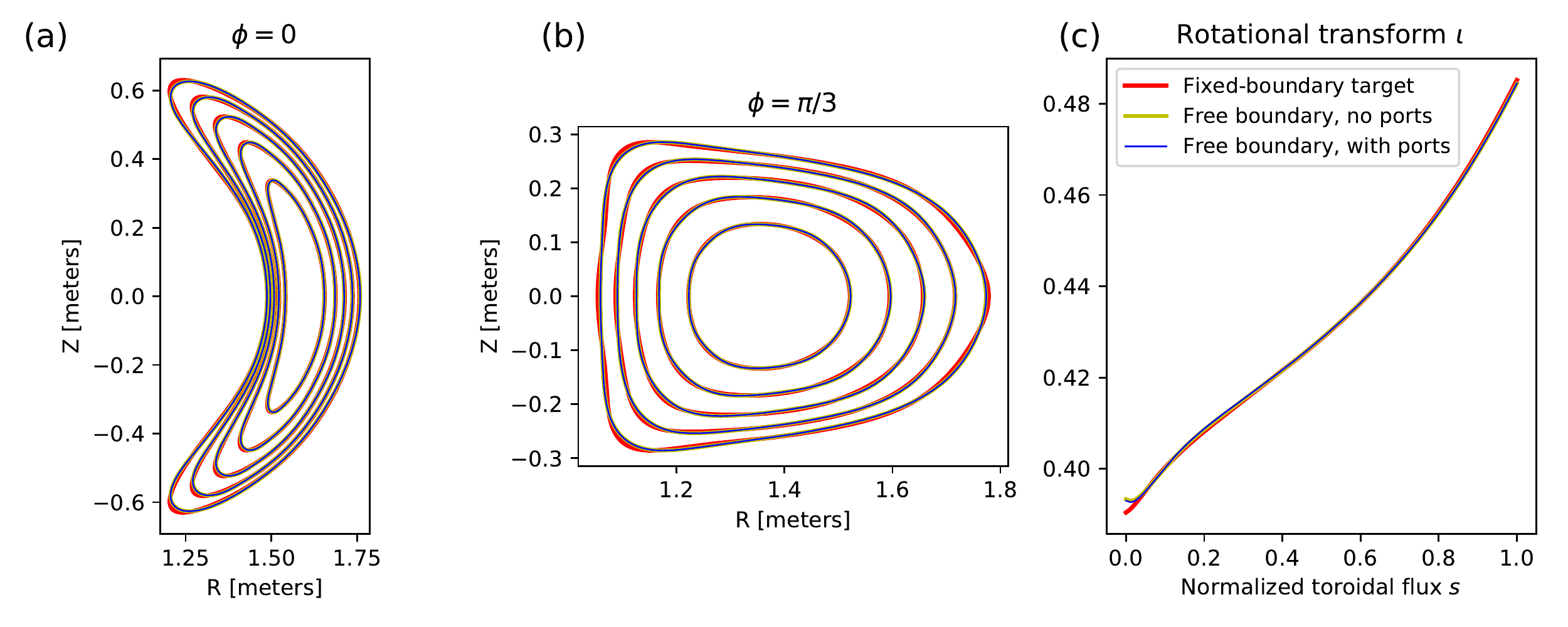}
  \caption{
The REGCOIL\_PM solutions reproduce the target flux surface shapes and rotational transform.
  }
\label{fig:freeBoundary}
\end{figure}

To evaluate whether a magnet design is adequate, it is necessary to compute the resulting free-boundary plasma configuration. To this end, figure \ref{fig:freeBoundary} shows a comparison of the original c09r00 target configuration with the configurations achieved with permanent magnets. To compute the latter, our REGCOIL\_PM implementation saves an MGRID file that is used as input to free-boundary VMEC \cite{VMEC1983,VMEC1986}. 
REGCOIL\_PM results are shown both with and without ports, corresponding to figures \ref{fig:XSecions_noPorts} and \ref{fig:xsections}.
Panels (a)-(b) of figure \ref{fig:freeBoundary} show that the magnetic axis and flux surface shapes achieved are very close to those of the target configuration. Panel (c) shows that the rotational transform profile is reproduced accurately as well.
Differences between the REGCOIL\_PM results with and without ports are barely perceptible, indicating again that it should be possible to include ports in the design.
More detailed analysis must be done to assess whether the small differences in flux surface shape have a meaningful effect on physics properties. Nonetheless, these preliminary results support the idea that producing the 0.5 T NCSX configuration with permanent magnets is feasible.

\subsection{Higher field}

\begin{figure}
  \centering
  \includegraphics[width=6.1in]{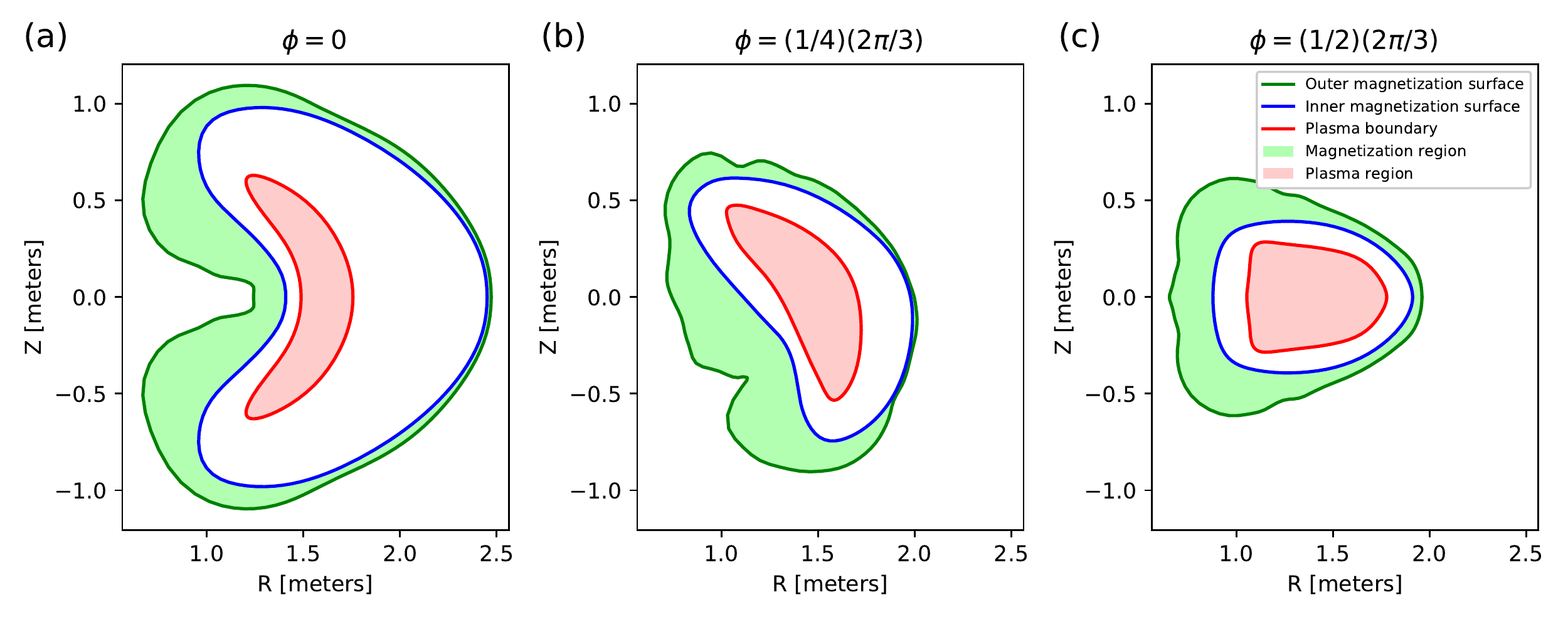}
  \caption{
If one attempts to raise the mean field magnitude of the NCSX example to 1 Tesla,
the magnet region becomes significantly thicker (compare to figure \ref{fig:XSecions_noPorts}).
  }
\label{fig:xsections_1T}
\end{figure}

Since the NCSX example developed in previous sections has a relatively weak magnetic field $\sim 0.5$ T, 
a natural question is whether the field magnitude can be increased. 
Here we examine the feasibility of doubling the field to 1 Tesla. In the approximation that the field produced by a permanent magnet is proportional to its thickness (figure \ref{fig:basicIdea}), doubling $\vect{B}$ would require a doubling of the magnet thickness. In fact the thickness must be more than doubled, since the new magnet that is introduced compared to the 0.5 Tesla case is farther from the plasma and so has less effect.
Figure \ref{fig:xsections_1T} shows the REGCOIL\_PM solution for the 1 Tesla case with no ports,
and a comparison to figure \ref{fig:XSecions_noPorts} makes clear that a significant increase in magnet thickness is indeed required. The magnetization volume for the 0.5 Tesla case is 2.0 m$^3$, compared to 4.9 m$^3$ for the 1 Tesla case. The magnet thickness for the 1 Tesla case is sufficiently large that the coordinate system in (\ref{eq:positionVector}) becomes singular, with $\sqrt{g}$ crossing zero. This issue is specific to the coordinate system we have chosen in regions where the inner surface is concave, and does not necessarily mean a 1 Tesla solution is impossible. However the significant volume occupied by the magnets in figure \ref{fig:xsections_1T} suggests that a $\ge 1$ Tesla NCSX with the existing toroidal field (TF) coils and permanent magnets is likely infeasible. It may well be possible to obtain 1 Telsa solutions if the TF coils were shifted or rotated, or if a different plasma geometry is chosen.

\subsection{Benchmark with FAMUS}

It is interesting to compare the results of
REGCOIL\_PM to the topology optimization method
described in \cite{ZhuTopology}. The latter approach is implemented in the code FAMUS.
In topology optimization, the presence or absence of a magnet at a given location is represented by a continuous variable $\rho \in [0, 1]$, and optimization is used to penalize intermediate values in the range $(0,1)$ so $\rho \approx 0$ or $1$ at most locations.
The FAMUS and REGCOIL\_PM approaches are expected to each have advantages and disadvantages.
The potential advantages of REGCOIL\_PM have already been described.
Topology optimization is more flexible with respect to the magnet geometry, with no restriction that all magnets have one fixed surface specified by the user.

We carry out a comparison between the two codes
for the 0.5 T NCSX
case with no ports or plasma current.
We first obtain a FAMUS solution,
considering dipoles allowed to lie within 14 cm of the NCSX vacuum vessel in the direction away from the plasma. The grid of allowed dipole locations has a resolution of 14 points radially, 64 points in $\theta$, and 384 points in $\phi$ (considering all field periods). The level of regularization in FAMUS is set by hand to achieve a plausible solution, with $f_B = 2.12\times 10^{-6}$ T$^2$ m$^2$.
Then $\lambda$ in REGCOIL\_PM is adjusted to match this value of $f_B$, with the result $\lambda = 8.51\times 10^{-16}$ T$^2 /$ A$^2$. Both codes achieve the same target magnetization $M_t = 1.1 \times 10^6$ A$/$m$^2$.
An effective volume of the permanent magnet region can be defined in FAMUS by $\sum_j |\vect{m}_j| / M_t$ where $\vect{m}_j$ are the discrete dipole moments; the result for this case is 2.32 m$^3$. The permanent magnet volume of the REGCOIL\_PM solution is slightly lower, 1.96 m$^3$.

\begin{figure}
  \centering
  \includegraphics[width=6.1in]{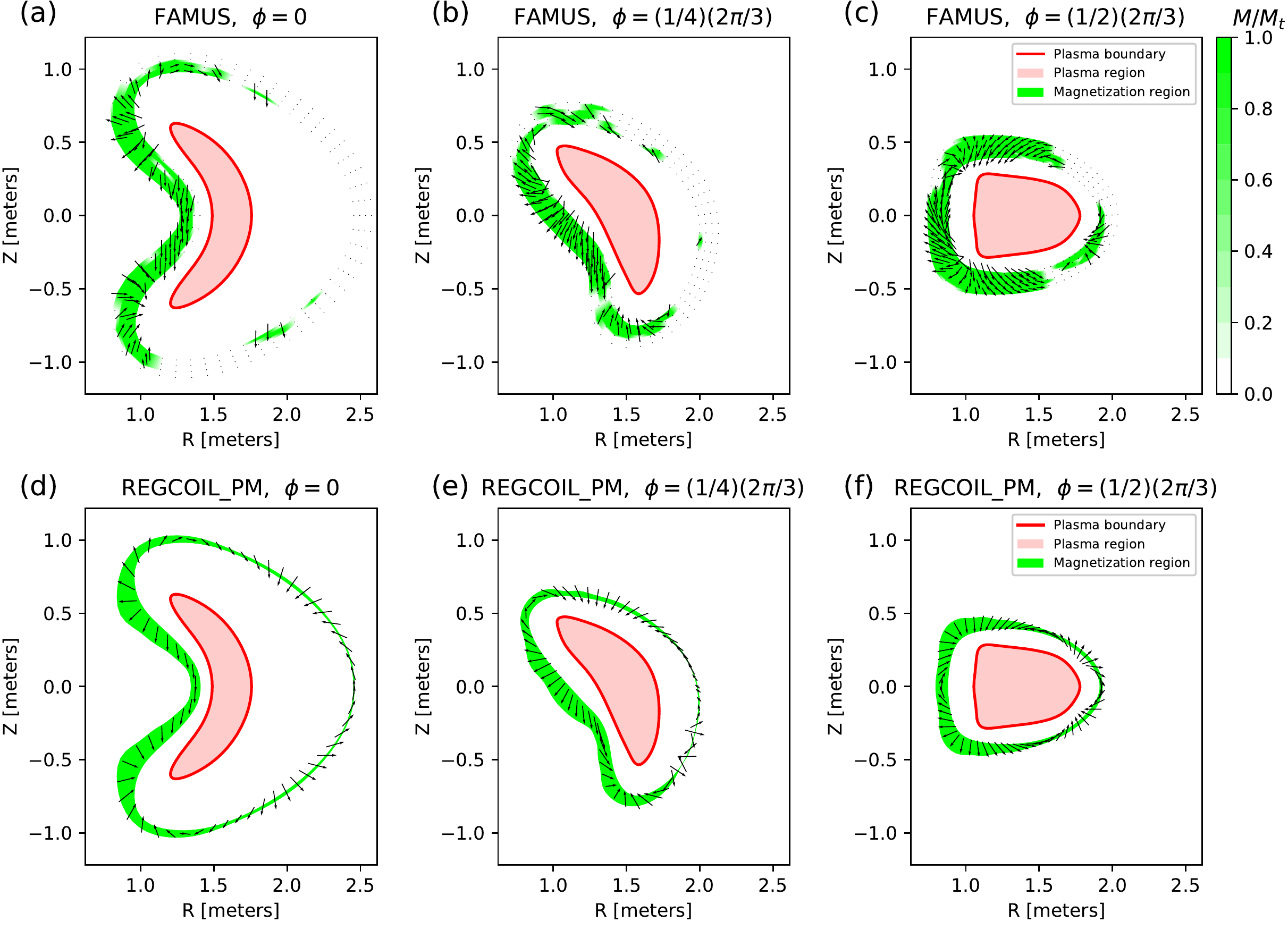}
  \caption{
Comparison between REGCOIL\_PM and the topology optimization code FAMUS at matched $f_B$. Black arrows indicate the direction of the magnetization.
  }
\label{fig:famus}
\end{figure}

The results of the two codes are shown in figure \ref{fig:famus}. 
It can be seen that $\rho = 0$ or 1 nearly everywhere in the FAMUS solution.
For both codes, black arrows display the magnetization vector's projection into the $(R,Z)$-plane. While $|\vect{M}|$ is exactly uniform in the REGCOIL\_PM solution and very nearly uniform in the FAMUS solution, the arrow lengths vary since a $\phi$ component may be present. In panels (a)-(c), the $\vect{M}$ vectors are shown for 4 of the 14 radial grid locations in FAMUS. In REGCOIL\_PM, where there is no radial variation, only a single arrow is shown. 
The dipole locations in FAMUS are shifted from the symmetry planes by half of the grid spacing (i.e. by $2\pi/768$ radians), so the figures show the nearest planes of dipoles to the given $\phi$.
There are many similarities between the solutions from the two codes. Both codes yield a thicker magnet layer on the small-$R$ side of the plasma. In these thick regions, the direction of the magnetization is very similar between the two codes. 
At the large-$R$ side, the REGCOIL\_PM solution has a thin magnet layer, whereas FAMUS eliminates the magnets in many of these regions. These two different magnet configurations both produce a small field error $f_B$, demonstrating again that there is significant flexibility in the magnet design.


\section{Conclusions}
\label{sec:conclusions}

In summary, we have demonstrated an algorithm for computing an arrangement of permanent magnets outside of a target volume that produces a desired spatially-dependent magnetic field inside the volume. While the algorithm is applied here to stellarators, the method could be used for other applications as well.
The method here results in a binary magnetization magnitude: at every point $M$ is either zero or equal to a target value $M_t$. This feature is advantageous since any volume occupied by magnetization of less than the maximum commercially available magnitude is an inefficient use of space. The method also does not place constraints on the direction of $\vect{M}$, meaning that Halbach solutions with rotating $\vect{M}$ are obtained automatically. While we have not rigorously proved stability or existence of a unique fixed point, the method appears to give a unique result independent of the initial guess, meaning users need not worry about how to choose a good initial condition.

In this work for expediency we have considered the case of magnets in a domain with smooth boundary and with smoothly varying $\vect{M}$.
This approximation is likely inaccurate for a serious experimental design. However it appears straightforward to extend the REGCOIL\_PM algorithm to a more realistic case of discrete magnet blocks with a uniform direction of $\vect{M}$ in each block. Each block $k$ would be parameterized with a thickness parameter $d_k$. The linear-least-squares solve would have three degrees of freedom per block, one for each coordinate of the block's $\vect{M}$ vector. The $d_k$ parameter of each block could be updated by applying the same fixed-point iteration used here to each block. This idea will be explored in future work.

Even without this extension to discrete magnet blocks, REGCOIL\_PM could be valuable as part of optimization of the plasma shape, i.e. the first stage in the standard two-stage stellarator design. At each iteration of the plasma optimization, REGCOIL\_PM could be called, and the resulting magnet thickness could be penalized in the objective function along with other physics quantities. One could thereby find plasma configurations that can be produced with a relatively low volume of permanent magnets. Inside this optimization, robustness and speed of a code are more important than detailed modeling of all engineering factors, and so the `smooth' REGCOIL\_PM of the present paper would be sufficient and well suited. For this application the number of fixed-point iterations (eq (\ref{eq:d_update})) could be very small, perhaps one, since the magnet thickness need not be precise. Or, the fixed-point iteration could be avoided entirely, and rather the peak magnitude $M$ from the linear-least-squares solution with uniform $d$ could be penalized.


\ack
Input from Steven Cowley, David Gates, Ken Hammond, Per Helander, 
Tonatiuh S\'{a}nchez-Vizuet, and
Michael Zarnstorff is gratefully acknowledged.
This work was supported by the
U.S. Department of Energy
under Contract No. DE-AC02-09CH11466.


\appendix


\section{Jacobian}
\label{apx:Jacobian}

Here we derive (\ref{eq:Jacobian}). Applying $\partial/\partial s$, $\partial/\partial \theta$, and $\partial/\partial \zeta$ to (\ref{eq:positionVector}), one finds
\begin{align}
\label{eq:JacobianStep1}
    \sqrt{g}=&\frac{\partial \vect{r}'}{\partial s} \cdot 
    \frac{\partial \vect{r}'}{\partial \theta}\times \frac{\partial \vect{r}'}{\partial \zeta}
    \\
    =&-\sigma d\left[N - \sigma s d \left( \vect{n}\cdot\frac{\partial\vect{r}_I}{\partial\theta}\times \frac{\partial \vect{n}}{\partial\zeta}+
    \vect{n}\cdot\frac{\partial\vect{n}}{\partial\theta}\times \frac{\partial \vect{r}_I}{\partial\zeta}\right)
    -s^2 d^2 \vect{n}\cdot \frac{\partial\vect{n}}{\partial\theta}\times\frac{\partial\vect{n}}{\partial\zeta}\right].
    \nonumber
\end{align}
The quantity in parentheses is $2HN$, as shown in appendix A of \cite{LandremanPaul2018}. The last term in (\ref{eq:JacobianStep1}) is evaluated by differentiating $\vect{n}=\vect{N}/N$ with (\ref{eq:Ndef}) to obtain
\begin{align}
    \frac{\partial\vect{n}}{\partial\theta}
    =\frac{1}{N}\left[ \frac{\partial^2 \vect{r}_I}{\partial\theta\partial\zeta}\times\frac{\partial\vect{r}_I}{\partial\theta}
    +\frac{\partial\vect{r}_I}{\partial\zeta}\times\frac{\partial^2\vect{r}_I}{\partial\theta^2} 
    - \vect{n}\frac{\partial N}{\partial\theta}\right],
    \\
    \frac{\partial\vect{n}}{\partial\zeta}
    =\frac{1}{N}\left[ \frac{\partial^2 \vect{r}_I}{\partial\zeta^2}\times\frac{\partial\vect{r}_I}{\partial\theta}
    +\frac{\partial\vect{r}_I}{\partial\zeta}\times\frac{\partial^2\vect{r}_I}{\partial\theta\partial\zeta} 
    - \vect{n}\frac{\partial N}{\partial\zeta}\right].
\end{align}
Straightforward manipulation then gives (\ref{eq:Jacobian}).


\section{Integrals for section \ref{sec:benchmark}}
\label{apx:integrals}
   
Here we derive expressions
(\ref{eq:potentialUniform}) and (\ref{eq:potentialNormal}).
We start by inserting (\ref{eq:uniformMagnitudeDipoles}) (for uniform-magnitude dipoles) or (\ref{eq:normalDipoles}) (for dipoles normal to the magnet surface) into
(\ref{eq:generalPotential}). The results are
\begin{equation}
    \Phi_{uni} = -\frac{\mu_0 \eta b \hat{m}}{4\pi} \int_0^{2\pi} d\theta' \int_{-\infty}^{\infty}dz'
    \frac{(r\cos\theta - b \cos\theta')\cos n\theta' + (r\sin\theta - b\sin\theta')\sin n\theta'}
    {[(r\cos\theta-b\cos\theta')^2+(r\sin\theta-b\sin\theta')^2+(z-z')^2]^{3/2}}
\end{equation}
and
\begin{equation}
    \Phi_{nor} = -\frac{\mu_0 \eta b \bar{m}}{4\pi} \int_0^{2\pi} d\theta' \int_{-\infty}^{\infty}dz'
    \frac{[(r\cos\theta - b \cos\theta')\cos \theta' + (r\sin\theta - b\sin\theta')\sin \theta']\cos\ell\theta'}
    {[(r\cos\theta-b\cos\theta')^2+(r\sin\theta-b\sin\theta')^2+(z-z')^2]^{3/2}}
\end{equation}
respectively. The $z'$ integrals are evaluated using
$\int_{-\infty}^{\infty}dz' [q+(z-z')^2]^{-3/2}=2/q$ for $q>0$.
Changing the remaining integration variable to $\gamma=\theta'-\theta$
and using an angle-sum trigonometric identity, one finds
\begin{equation}
    \Phi_{uni}=-\frac{\mu_0 \eta \hat{m}}{2\pi}
    \int_0^{2\pi}d\gamma
    \frac{A}
    {1+\frac{r^2}{b^2}-2\frac{r}{b}\cos\gamma}
\end{equation}
with
\begin{align}
    A=&\frac{r}{b}[\cos n\gamma \cos((n-1)\theta) - \sin n\gamma \sin((n-1)\theta)] \\ 
    & -\cos((n-1)\gamma)\cos((n-1)\theta) + \sin((n-1)\gamma)\sin((n-1)\theta)\nonumber
\end{align}
and
\begin{equation}
    \Phi_{nor}=-\frac{\mu_0 \eta \bar{m}}{2\pi}
    \int_0^{2\pi}d\gamma
    \frac{\left[\frac{r}{b}\cos\gamma-1\right][\cos\ell\gamma \cos\ell\theta - \sin\ell\gamma \sin\ell\theta]}
    {1+\frac{r^2}{b^2}-2\frac{r}{b}\cos\gamma}.
\end{equation}
The contributions from terms $\propto \sin\ell\gamma$, $\sin n\gamma$, and $\sin((n-1)\gamma)$ all vanish. The remaining integrals can be evaluated using
\begin{align}
    \int_0^{\pi}d\gamma \frac{\cos n\gamma}{1-2\rho\cos\gamma+\rho^2}
    =&\frac{\pi \rho^n}{1-\rho^2} \;\;\; \mbox{for} \;\;
    \rho^2<1, \; n\ge 0,
\\
    \int_0^{\pi}d\gamma \frac{\cos n\gamma \cos\gamma}{1-2\rho\cos\gamma+\rho^2}
    =& \left\{ 
    \begin{array}{ll}
    \frac{\pi}{2} \frac{1+\rho^2}{1-\rho^2}\rho^{n-1} & \mbox{for}\; \rho^2<1, \; n \ge 1, \\
    \frac{\pi \rho}{1-\rho^2} & \mbox{for}\; \rho^2<1, \; n=0.
    \end{array}
    \right.
\end{align}
The results are 
(\ref{eq:potentialUniform}) and (\ref{eq:potentialNormal}).

\section*{References}

\bibliographystyle{unsrt}
\bibliography{regcoil_pm}

\end{document}